\documentclass[prd,nobibnotes,showpacs,twocolumn]{revtex4}
\usepackage{graphics,color,array,dcolumn}
\usepackage{calc}

\newcommand{\be}{\begin{equation}}
\newcommand{\ee}{\end{equation}}
\newcommand{\Qa}[2]{\mathrm{Q_1}\! \left[\frac{#1}{#2}\right]}
\newcommand{\Qb}[2]{\mathrm{Q_2}\! \left[\frac{#1}{#2}\right]}
\newcommand{\Qabis}[1]{\mathrm{Q_1}\! \left[#1}
\newcommand{\Qbbis}[1]{\mathrm{Q_2}\! \left[#1}
\renewcommand{\P}{{\mathcal{P}}}
\newcommand{\Pt}{{\partial_t\P}}
\newcommand{\R}{{\mathcal{R}}}

\newcommand{\x}[1]{\xi_{#1}}
\renewcommand{\l}[1]{\lambda_{#1}}
\usepackage{amsmath}
\usepackage{amssymb}
\usepackage{xspace}

\newcommand{\lambdah}{{\lambda}}
\newcommand{\gh}{{g}}

\newlength{\figurewidth}
\allowdisplaybreaks[1]

\begin{document}
\setlength{\figurewidth}{\columnwidth}

\title{Asymptotic Safety of Gravity Coupled to Matter}
\author{Roberto Percacci}
\email{percacci@sissa.it}
\author{Daniele Perini}
\email{perini@he.sissa.it}
\affiliation{SISSA, via Beirut 4, I-34014 Trieste, Italy}
\affiliation{INFN, Sezione di Trieste, Italy}


\begin{abstract}
Nonperturbative treatments of the UV limit of pure gravity
suggest that it admits a stable fixed point with positive
Newton's constant and cosmological constant.
We prove that this result is stable under the addition of 
a scalar field with a generic potential
and nonminimal couplings to the scalar curvature.
There is a fixed point where the mass 
and all nonminimal scalar interactions vanish while 
the gravitational couplings have values which are almost 
identical to the pure gravity case.
We discuss the linearized flow around this fixed point
and find that the critical surface is four-dimensional.
In the presence of other, arbitrary, massless minimally 
coupled matter fields, the existence of the fixed point,
the sign of the cosmological constant and the dimension of the
critical surface depend on the type and number of fields. 
In particular, for some matter content,
there exist polynomial asymptotically
free scalar potentials, thus providing a 
solution to the well-known problem of triviality.

\end{abstract} \maketitle


\section{Introduction}

The failure of perturbative approaches to quantum gravity does not
necessarily imply that quantum gravity does not exist as a field
theory. There is still in principle the possibility that the theory
could be ``nonperturbatively quantized''.
To understand what this means, one has to look at the
Renormalization Group (RG), \emph{i.\ e.}\ the flow of the coupling
constants $\tilde g_i(k)$ as a certain external momentum parameter $k$ is changed%
\footnote{the precise physical meaning of the parameter $k$ depends
on the specific problem that one is addressing; it is usually
the momentum of some particle entering into the process under
study, or the inverse of some characteristic lenght of the system.
In general, $k$ has the meaning of an IR cutoff, because the
effective theory describing the process must include the effect of
all the fluctuations of the fields with momenta larger than $k$.}.
It is customary to take $k$ as unit of mass; if $\tilde g_i$ has
dimension $d_i$ in units of mass, we define
dimensionless couplings $g_i=\tilde g_i k^{-d_i}$.
The RG flow is then given by the integral curves of a vectorfield $\beta$
in the space of all couplings, whose components 
$\beta_i(g)=\partial_t g_i$ (with $t=\ln k$)
are the beta functions.
A Fixed Point (FP) is a point $g_*$ in the space of all couplings where 
\begin{equation}
\label{eq:fp}
\beta_i(g_*)=0\ .
\end{equation}
For example, in ordinary quantum field theories in Minkowski space,
the point where all couplings vanish is always a FP
(called the Gaussian FP),
because a free field theory does not have quantum corrections.

Suppose that the theory admits a FP.
One defines the critical surface to be the locus of points
that, under the RG evolution, are attracted towards the FP for $t\to\infty$. 
Starting from any point on the critical surface, 
the UV limit can be taken safely, because the couplings,
and as a consequence the physical reaction rates,
will be drawn towards the FP and hence remain finite
\footnote{In general in the action there will be couplings
whose values do not affect cross-sections and reaction rates. 
They are called inessential couplings.
Our reasoning applies only to the essential couplings.}
On the other hand, if one starts from a point not belonging to the 
critical surface, the RG evolution will generally lead to
divergences.
If the critical surface has finite dimension $c$, 
the theory will be predictive, because only $c-1$ parameters
will be left undetermined and
will have to be fixed by experiment at a given energy scale
(the last remaining parameter being the scale itself).
The special case when the FP with a finite dimensional critical 
surface is the Gaussian FP is equivalent to
the usual perturbative notion of renormalizability 
and asymptotic freedom.

This scenario for nonperturbative renormalizability has been discussed
specifically in a gravitational context in \cite{Weinberg:1979} where
this good property was called ``asymptotic safety''.  At the time some
encouraging results were obtained by studying gravity in $2+\epsilon$
dimensions \cite{Weinberg:1979,Gastmans:1978ad}%
, but the program soon came to a
halt essentially for want of technical tools.  It now appears that the
right tool to tackle this problem is the Exact RG Equation (ERGE), in
one of several guises that have appeared in the literature in the last
decade
\cite{Polchinski:1984gv,Bagnuls:2000ae,
Wetterich:1993yh}.

The ERGE is a differential equation that determines the RG flow of the
action. It can be viewed as a set of infinitely many first order 
differential equations for infinitely many variables (the coupling constants),
and therefore cannot generally be solved in practice.
A method that is commonly used to calculate nonperturbative beta functions is to
make a physically motivated Ansatz for the running effective action, 
typically containing a finite number of parameters, and insert this Ansatz
into the ERGE.

The ERGE, in the specific form discussed in \cite{Wetterich:1993yh},
has been applied to Einstein's theory in \cite{Reuter:1998cp,Dou:1998fg},
where the beta functions for Newton's constant and for the
cosmological constant were derived.
It was later realized that these beta functions actually admit
a nontrivial UV-attractive FP \cite{Souma:1999at}. 
The properties of this FP were further discussed in greater detail
in \cite{Lauscher:2001ya}. 
A particularly important issue is to prove that the FP 
is not an artifact of the Ansatz but is a genuine property of gravitation.
Several facts seem to indicate that the FP is quite robust.
It has been shown to exists in four spacetime dimensions for many 
different shapes of the cutoff, whereas in other dimensions
it only exists for certain cutoffs but not others.
Its properties have been shown to be only weakly dependent on the
shape of the cutoff, indicating that the truncation is self-consistent
\cite{Lauscher:2001rz}
.
It has been found in certain dimensionally-reduced versions
of the theory \cite{Forgacs:2002hz}.
It is also remarkable that Newton's constant always turns out to be
positive, a fact that could not in any way be guaranteed by
the general form of the equations.

The most important test, however, is the stability of the FP
against the addition of new couplings.
Each time we consider a new coupling,
whether remaining in the context of pure gravity
or if we introduce matter fields,
a new beta function has to vanish and therefore a new constraint 
has to be satisfied by the set of all couplings at the FP.  
It is therefore nontrivial that the FP still exists when we take 
into account additional couplings.

In the context of pure gravity, an important progress was made in
\cite{Lauscher:2002mb}, where it was shown
that the addition of a term quadratic in curvature does not spoil the
existence of the FP. In fact it turned out that the values of the
cosmological and Newton's constants at the FP are almost unaffected by
the new interaction, while the new coupling constant is quite small at
the FP.  This is far from being conclusive evidence, but it is
nevertheless an important result,
especially in view of the fact that another FP that was present in the
truncation with only two couplings---the Gaussian FP---does not exist
in the three coupling truncation.

As far as matter is concerned, we have recently considered the effect 
of minimally coupled, massless quantum fields of arbitrary spin \cite{Percacci:2002ie}. 
The only couplings taken into account were the
cosmological and Newton's constant, since the coefficients of the
matter kinetic terms can be normalized to their standard values by
field rescalings. It was shown that the existence of the FP,
the values of the cosmological constant and Newton's constant
at the FP and the dimension of the critical surface all depend on the 
type and number of fields present.  Altogether, the existence and
attractiveness of the FP puts some constraints on the number of matter
fields that are present.

In this paper we continue the analysis of coupled gravity and matter
systems and we begin to address the issue of matter couplings. 
There are many different couplings that are necessary to construct
realistic theories of the world, and we cannot possibly
take them all into account, so as a first step we shall consider
the simplest example, that of a self-interacting scalar field.
Aside from its role as a model for the Higgs field in unified theories,
a scalar field (the dilaton) appears in many popular theories of gravity.
It can therefore sometimes be regarded as part of the gravitational
sector, rather than the matter sector. This makes its properties
especially interesting in a gravitational context.
In this paper we shall not make any assumption about the physical
interpretation of the scalar field.

The class of actions that we consider is
\begin{equation}
\label{eq:class}
\Gamma[g,\phi]=\int\,\mathrm{d}^4x\sqrt{g}\left(V(\phi^2)- F(\phi^2)R
+{1\over2}g^{\mu\nu}\partial_\mu\phi\partial_\nu\phi\right)
\end{equation}
where the potential $V$ and the scalar-tensor coupling $F$
are arbitrary real analytic functions. (The RGEs for this system
have been studied earlier in \cite{Barvinsky:1993zg} 
Although it is not necessary for some of the results of this
paper, we shall assume that the potential has its minimum
at the origin. Then, we can identify
$F(0)=\kappa\equiv 1/16\pi G$, $G$ being Newton's constant, and 
$V(0)=2\kappa\Lambda$, where $\Lambda$
is the (dimension-two) cosmological constant.
It will appear that the behaviour of the couplings implicit
in the functions $V$ and $F$ is sufficiently systematic that
we are actually able to draw several conclusions involving an infinite
number of couplings.
For some purposes, however, we shall restrict our attention
to a five-parameter Ansatz, where $V$ is at most quartic in $\phi$
and $F$ is at most quadratic.

Aside from establishing the existence of a nontrivial FP,
the main question addressed in this paper will be the dimension 
of the critical surface.
In practice, this is done by linearizing the flow around the FP.
We define $v_i=g_i-g_{i*}$ the shift from the fixed point.
The linearized flow around the fixed point is described by the equations
\begin{equation}
\label{eq:linrg}
\partial_t v_i=M_{ij}v_j,
\end{equation}
where $M_{ij}={\partial \beta_i\over\partial g_j}$.
Let $P$  be the (generally complex) linear transformation that 
diagonalizes $M$:
$P^{-1}MP=diag(\alpha_1,\ldots,\alpha_N)$
(the columns of $P$ are the eigenvectors of $M$).
Defining $f=P^{-1}v$, one finds $\partial_t f_k=\alpha_k f_k$, so
$f_k(t)=e^{\alpha_k t}$.
Transforming back to the original variables,
the solution can be written 
$g_i(t)=g_{i*}+{\rm Re}(P_{ij} f_j(t))$.
It is easy to show that the eigenvalues $\alpha_i$ are
invariant under redefinitions of the couplings.

The eigenvalues with negative real part (which for brevity we shall call
the ``negative eigenvalues'') correspond to directions for which
the RG flow approaches the FP in the UV. 
The corresponding parameter $f_i$ is called a relevant parameter.
Those with positive real part (the ``positive eigenvalues'')
correspond to directions for which
the RG flow moves away from the FP in the UV.
The corresponding parameter $f_i$ is called an irrelevant parameter.
The parameters corresponding to purely imaginary eigenvalues are called marginal.
In the linearized theory, in order to approach the FP in the UV it is
therefore necessary to stay on the hyperplane spanned by the
eigenvectors with negative eigenvalues.
This hyperplane is the tangent space to the critical surface at the FP.
Therefore, the dimension of the critical surface is equal to
the number of negative eigenvalues.
On general grounds, one expects the critical surface to be finite dimensional.
In this way all but a finite number of couplings would be
fixed and the theory would be as predictive as a perturbatively
renormalizable theory.

We now give a brief summary of the results of this paper.
First of all, a nontrivial FP still exists with the Ansatz \eqref{eq:class}. 
The purely gravitational couplings (cosmological constant and Newton's constant,
which appear as the $\phi^2$-independent terms in the functions $V$ and $F$)
have the same values as in \cite{Percacci:2002ie}
and all the other couplings are equal to zero.
In a sense this is therefore ``the same'' fixed point that was considered
in \cite{Souma:1999at} and in \cite{Lauscher:2001ya,Reuter:2001ag}. 
It can also be regarded as a generalization of the Gaussian FP
of the pure scalar theory in flat space.
We therefore call it the Gaussian-Matter FP (GMFP).
We have performed a systematic search for other FP's
within a five-parameter truncation of the action, where $V$ and $F$
are polynomials containing at most terms of order $\phi^4$ and $\phi^2$
respectively.
Thus, in addition to the cosmological and Newton's
constant, we consider a scalar mass term, a quartic
self-interaction and a nonminimal coupling of the scalar field
to the scalar curvature.
Detailed numerical analyses have convinced us that there are no FP's
with nonzero scalar mass and couplings, for values of the
cosmological and Newton constant close to the ones of the pure-gravity FP.

Comparing to the results of the pure scalar theory,
the main effect of the coupling to gravity is to change the exponents
$\alpha_i$.
It turns out that of the two canonically marginal couplings, 
the $\phi^4$ coupling becomes irrelevant while the $\phi^2 R$
coupling becomes relevant. The other couplings preserve the character that
is implied by their canonical dimension; the critical dimension would thus 
be equal to four.

We then look at the effect of other matter fields on the FP.
The results of this investigation generalize those already reported
in \cite{Percacci:2002ie}. 
The behaviour of the beta functions is determined by two parameters
that depend on the number of fields, and the existence of the FP
depends on the values of these parameters.
In this way the existence of the FP yields constraints on the type 
and number of matter fields. These constraints appear to be satisfied 
by popular unified models.
The existence region is subdivided into subregions with varying numbers
of attractive directions. 
In the region that we have explored, comprising large numbers of matter
fields, the dimension of the critical surface is always finite.
In particular, there are regions in which the attractive directions correspond to
nontrivial polynomial potentials of degree four or higher.
This yields a neat solution of a long-standing puzzle.
In a pure scalar theory in flat space, the Gaussian FP is IR attractive
(all couplings are irrelevant). As a consequence, when one takes the
continuum limit at the Gaussian FP, the renormalized theory is free.
This result is not an obstacle in the context of an effective field theory,
but it has to be somehow circumvented if scalar fields have to appear in
a fundamental field theory. The coupling to gravity is a natural context
for a solution of this issue. Our results imply that there exist theories of gravity
coupled to matter such that the renormalized scalar potential has finitely
many nonzero couplings in the continuum limit.
This seems to indicate that the interaction with gravity 
(and, indirectly, with the other matter fields) solves the problem of the 
triviality of the scalar theory.

This paper is organized as follows.
In Section II, by way of introduction, we derive the ERGE and we 
use it to prove some well-known
results on the Gaussian FP of a pure scalar theory in flat space.
In Section III we consider the modifications of the beta functions
due to gravity, we prove the existence of the Gaussian-Matter FP (GMFP) and
we discuss the (negative) results of the numerical search 
for other FP's.
Section IV is devoted to the properties of the GMFP.
We analyze the linearized flow around the
GMFP and show that the coupling to gravity affects the
dimensions of the couplings, shifting them relative to the
canonical values.
In Section V we consider the effect of other massless,
minimally coupled matter fields on the GMFP. 
In Section VI we will consider in some detail the
dependence of our results on the shape of
the cutoff function and on the gauge--fixing parameter.
Finally in Section VII we make some concluding remarks.

All results are derived in the case of Euclidean signature,
in four dimensions.
Since the expressions of the beta functions are extremely lengthy,
in deriving our results
we have made extensive use of algebraic manipulation software.
\section{The Gaussian FP in pure scalar theory}
We begin by considering the case of a single scalar field without
gravity and a generic even potential
\begin{equation}
\label{eq:scalar_potential}
V(\phi)=\sum_{n=0}^\infty \tilde\lambda_{2n}\phi^{2n}.
\end{equation}
In this section we assume that 
$V(0)=\tilde\lambda_0=2\Lambda\kappa=0$;
the first nonzero term is the mass $\tilde\lambda_2={1\over2}m^2$,
while $\tilde\lambda_4$ is the usual quartic coupling.
The couplings $\tilde\lambda_{2n}$ have dimension $mass^{2(2-n)}$,
so the usual power-counting arguments tell us that the terms
in \eqref{eq:scalar_potential} with $n>2$ are perturbatively 
nonrenormalizable, while the term $n=2$
is marginal. We will now rederive this result within the
formalism of the ERGEs. This will set the stage for
further developments in later sections.

To derive the ERGE, one begins by modifying the classical propagator
by adding to the action a term quadratic in the fields which in
momentum space can be written $\Delta S_k(\phi)=\frac{1}{2}\int d^4q
\phi(-q) R_k(z) \phi(q)$, where $z=q^2$.  The effect of this term must
be to suppress the propagation of field modes with momenta smaller
than $k$, while leaving the modes with momenta larger than $k$
unaffected.  This is the case if the smooth cutoff function $R_k$ is
chosen to tend to zero for $z\gg k^2$ and to a constant for $z\to 0$.
For numerical work, in this paper we will work with cutoffs of the
form
\begin{equation}
\label{eq:cutoff}
R_k(z)={2a z e^{-2az/k^2}\over 1-e^{-2az/k^2}}\ ,
\end{equation}
with $a$ a free parameter.
We then define a scale--dependent generating functional
of connected Green functions
$$
W_k[J]=-\mathrm{ln}\int (D\phi)\mathrm{exp}(-(S+\Delta S_k+\int J\phi))
$$
such that $\frac{\delta W_k}{\delta J}\Big|_{J=0}=\langle \phi \rangle$
and a scale-dependent effective action
$\Gamma_k[\phi_{cl}]=\tilde \Gamma_k[\phi_{cl}]-\Delta S_k[\phi_{cl}]$, 
where $\tilde \Gamma_k[\phi_{cl}]=W_k[J]-\int J\phi_{cl}$
is obtained from $W_k[J]$ by the usual Legendre-transform procedure.
The scale-dependent effective action tends to the bare action $S$
when $k$ tends to the UV cutoff, and to the ordinary effective
action for $k\to 0$.
We have
\begin{equation}
\partial_t W_k=\partial_t \langle\Delta S_k\rangle
=\frac12\mathrm{Tr}\langle\phi\phi\rangle\partial_t R_k,
\end{equation}
where the trace is over all Fourier modes (and internal indices, if there
were any).
Then,
\begin{align}
\partial_t \Gamma_k[\phi_{\mathrm cl}]=&
\partial_t W_k[J]-\partial_t \Delta S_k[\phi_{\mathrm cl}]=\notag\\
&=\frac12\mathrm{Tr}(\langle\phi\phi\rangle-\langle\phi\rangle\langle\phi\rangle)\partial_t R_k\notag\\
&=-\frac12\mathrm{Tr}\frac{\delta^2 W_k}{\delta J\delta J}\partial_t R_k.
\end{align}
Applying the standard identity
\begin{equation}
\frac{\delta^2 W_k}{\delta J\delta J}=
-\left(\frac{\delta^2\tilde\Gamma_k}{\delta\phi_{cl}\delta\phi_{cl}}\right)^{-1},
\end{equation}
one then obtains the ERGE \cite{Wetterich:2001kr}
\begin{equation}
\label{eq:ERGE}
\partial_t\Gamma_k=
{1\over2}\mathrm{Tr}\left({\delta^2\Gamma_k\over\delta\phi\delta\phi}+R_k\right)^{-1}
\partial_tR_k.
\end{equation}
In the previous formula and in the following we shall drop the subscript in $\phi_{cl}$;
this should not cause any confusion.
The ERGE describes the flow of the functional $\Gamma_k$ with the scale $k$.
In order to extract beta functions, one has to resort to approximations.
A common procedure is to make an Ansatz about the form of $\Gamma_k$
and to insert it into the ERGE. 
Of course the beta functions obtained in this way are no longer exact:
one loses all information about the dependence of the beta functions 
on the parameters that have been left out of the Ansatz.
Nevertheless, the results do contain information that is not accessible
in perturbation theory and they have been shown to yield numerically 
accurate values in many circumstances \cite{Bagnuls:2000ae,Bonanno:2000sy}.
We now apply this procedure to the scalar theory.

Introducing in equation \eqref{eq:ERGE} the truncation
\mbox{$\Gamma_k(\phi)=\int d^4x \left[-\frac12\phi\partial^2\phi+V(\phi^2)\right]$},
where $V$ is a $k$-dependent potential, gives
\begin{equation}
\partial_t\Gamma_k=
{1\over2}\mathrm{Tr}\left(
{\partial_tP_k\over
P_k+V'+4\phi^2V''}
\right).
\end{equation}
where a prime denotes the derivative with respect to $\phi^2$,
the trace can be understood as an integration over momenta.
It can be reexpressed as:

\begin{equation}
\partial_t\Gamma_k=
{\mathit{Vol}\over 32\pi^2}\tilde Q_2\left({\partial_tP_k\over
P_k+V'+4\phi^2V''}\right)
\end{equation}
where $\mathit{Vol}=\int d^4x$ denotes the volume of spacetime and
\begin{equation}
\tilde Q_n[f]=\frac1{\Gamma(n)}\int_0^{+\infty}\mathrm{d}z\,z^{n-1}f(z).
\end{equation}
The coupling constants can be extracted from the potential by
\begin{equation}
\tilde\lambda_{2n}=
{1\over n!}{\partial^n V\over\partial (\phi^2)^n}\Big|_{\phi=0}\ .
\end{equation}
In order to look for a fixed point one has to
define dimensionless couplings
$\lambda_{2n}=k^{2(n-2)}\tilde\lambda_{2n}$.
The corresponding beta functions are given by
\begin{equation}
\partial_t \lambda_{2n}=
2(n-2)\lambda_{2n}+{k^{2(n-2)}\over Vol}{1\over n!}
{\partial^n\over\partial (\phi^2)^n}\partial_t \Gamma_k \Big|_{\phi=0}.
\end{equation}
Explicitly, the first few beta functions are given by
\begin{widetext} 
\begin{subequations}
\label{eq:scalar.beta.functions}
\begin{align}
\label{eq:beta.lambda2} 
\partial_t\l2=&
-2\l2-{12\l4\over 32\pi^2}
Q_2\left({\partial_t\P\over \P+2\l2}
\right),\\
\label{eq:beta.lambda4} 
\partial_t\l4=&
{1\over 32\pi^2}\left[-30\l6
Q_2\left({\partial_t\P\over (\P+2\l2)^2}\right)+144\l4^2 
Q_2\left({\partial_t\P\over (\P+2\l2)^3}\right)\right],\\
\label{eq:beta.lambda6}
\partial_t\l6=&
2\l6
+{1\over 32\pi^2}\left[
-56\l8
Q_2\left({\partial_t\P\over (\P+2\l2)^2}\right)
+720\l4\l6 
Q_2\left({\partial_t\P\over (\P+2\l2)^3}\right)
-1728\l4^3 
Q_2\left({\partial_t\P\over (\P+2\l2)^4}\right)
\right],\\
\label{eq:beta.lambda8}
\partial_t\l8=&
4\l8
+{1\over 32\pi^2}\Bigg[
-90\lambda_{10}
Q_2\left({\partial_t\P\over (\P+2\l2)^2}\right)
+1344\l4\l8 
Q_2\left({\partial_t\P\over (\P+2\l2)^3}\right)\notag\\
&
+900\l6^2 
Q_2\left({\partial_t\P\over (\P+2\l2)^3}\right)
-8640\l6\l4^2 
Q_2\left({\partial_t\P\over (\P+2\l2)^4}\right)
+20736\l4^4 
Q_2\left({\partial_t\P\over (\P+2\l2)^5}\right)
\Bigg],
\end{align}
\end{subequations} 
\end{widetext}
where $Q_n[f]=k^{-2n}\tilde Q_n[f]$ is a dimensionless integral,
$\R=k^{-2}R_k$ is a dimensionless cutoff and
and $\P=k^{-2}P_k$ is a dimensionless modified propagator. 

This theory admits a well-known Gaussian FP: if we set
$\lambda_2=0$, equation \eqref{eq:beta.lambda2} implies
$\lambda_4=0$, equation \eqref{eq:beta.lambda4} then implies
$\lambda_6=0$, and so on: recursively all couplings are found to
be zero.  
This is not the only solution of the coupled system. 
One can fix an arbitrary value of $\lambda_2$
and the equations then recursively determine all the other couplings
\cite{Halpern:1995vw}
. 
However, when $\l2\not= 0$ these potentials become singular at a 
finite value of $\phi$
and therefore are not considered to be physically acceptable
\cite{Morris:1996nx}
.
In what follows we will restrict our attention to the Gaussian FP.

We now study the critical surface in the neighborhood of the FP
using the linearized RG equation \eqref{eq:linrg}.
Let $\beta_{2n}=\partial_t\lambda_{2n}$ and let $M_{ij}={\partial
\beta_{2i}\over\partial\lambda_{2j}}$.
It appears from equations \eqref{eq:scalar.beta.functions} 
(as well as from dimensional and diagrammatic
considerations) that the $2n$-th beta function is a polynomial in the
couplings $\lambda_4,\ldots,\lambda_{2n+2}$, linear in
$\lambda_{2n+2}$. Therefore the elements of the matrix $M_{ij}$
with $j>i+1$ are zero. On the other hand, since all $\lambda_{2n}$
are zero at the FP, when the derivatives are evaluated at the FP only
the terms linear in the couplings remain.  These are exactly the terms
on the diagonal, which are equal to the canonical dimensions of the
couplings, and the terms on the second diagonal ($i=j+1$), which are
equal to
\begin{equation}
M_{i\;i+1}={\partial\beta_{2n}\over\partial\lambda_{2n+2}}=
(2n+1)(n+1) c
\end{equation}
where 
$c=-{1\over 16\pi^2}Q_2\left({\Pt\over\P^2}\right)$
All other terms are zero.
Numerically, the integral
$Q_2\left({\Pt\over\P^2}\right)$
is equal to 0.924 for $a=2$.

Therefore, the matrix $M$ has the following form:
\begin{equation}
\label{eq:pure.scalar.matrix}
\left(
\begin{array}{c c c c c}
-2&
6c&
0&0&\ldots\\
0&0&
15c&
0&\ldots\\
0&0&2&
28c&
\ldots\\
0&0&0&4&\ldots\cr
\ldots&\ldots&\ldots&\ldots&\ldots
\end{array}
\right).
\end{equation}
The eigenvalue problem for this infinite matrix yields the
recursion relation
\begin{equation}
\label{eq:recursion}
\l{2n+2}={2(n-2)-\mu\over (2n+1)(n+1)c}\l{2n}
\end{equation}
where $\mu$ is the eigenvalue.
This relation can have two types of solutions.
If we assume that the potential is a finite polynomial
of order $K$, Eq.\eqref{eq:recursion} implies that $\mu=2(K-2)$.
These eigenvalues are just the diagonal elements of the matrix 
\eqref{eq:pure.scalar.matrix}.
The corresponding eigenvectors are the columns of the following matrix $P$:
\begin{equation}
\label{eq:pure.scalar.eigenvectors}
\left(
\begin{array}{c c c c c}
1&
-0.0175512&
3.84804\times 10^{-3}&
1.04825\times 10^{-5}&
\ldots\\
0&
0.999846&
-0.0438425&
1.79148\times 10^{-3}&
\ldots\\
0&
0&
0.999038&
-0.0816446&
\ldots\\
0&
0&
0&
0.99666&
\ldots\cr
\ldots&\ldots&\ldots&\ldots&\ldots
\end{array}
\right).
\end{equation}
The eigenvalues are equal to the canonical
dimensions of the couplings, so that the relevant, irrelevant and
marginal couplings correspond exactly to the couplings that are
superrenormalizable, nonrenormalizable and renormalizable 
in the perturbative sense.

These polynomial potentials suffer from the well-known problem
of triviality. 
Consider the scalar theory regularized with a UV cutoff 
$\Lambda_{UV}$ and the IR cutoff $k$.
Keeping $k$ fixed and letting $\Lambda_{UV}\to\infty$
(the continuum limit)
has the same effect as keeping $\Lambda_{UV}$ fixed and
letting $k\to 0$.
An irrelevant coupling tends to zero for $k\to 0$, 
and therefore, for any fixed $k$
it will tend to zero in the continuum limit.
This will be the case for all $\l{2i}$ with $i\geq 2$,
so the theory is non-interacting in the continuum limit.
(Our analysis only says that the couplings from $\l6$ upwards
have to be zero; the hard part is to prove that also the
marginal coupling $\l4$ tends to zero. For this, one has
to go beyond the linearized analysis
\cite{Fernandez:1992jh}.)

There is also another type of eigenvectors, corresponding to
nonpolynomial potentials, that avoids the problem of triviality. 
If we do not assume that $\l{2K+2}=0$ for some
$K$, the recursion relation \eqref{eq:recursion}
can be solved for the $\l{2n}$ in terms of the free parameters $\l2$ and
$\mu$, yielding a potential that can be written as a Kummer function
\cite{Halpern:1995vw}.
There are (negative) values of $\mu$ for which the potential has all the physically
desirable properties (positivity at $\infty$, symmetry breaking).
They are therefore nontrivial asymptotically free scalar theories.
However, there are infinitely many attractive directions and
therefore these theories do not satisfy the conditions for
asymptotic safety.

This concludes our brief review of the ERGE for a scalar field theory.

\section{The coupled system}
We now consider the coupling of the scalar theory to gravity,
using the Ansatz \eqref{eq:class} for the running effective action.
This will obviously change the beta functions of the scalar potential;  
in addition we will have to take into account also
the beta functions of the gravitational couplings.
These are given by the Taylor expansion coefficients of the 
function $F(\phi^2)$ of equation \eqref{eq:class},
which we write as follows:
\begin{equation}
F(\phi^2)=\sum_{n=0}^\infty \tilde\xi_{2n}\phi^{2n}.
\end{equation}
The first term in the expansion can be identified with the
(inverse) Newton constant: $\xi_0=\kappa=1/(16\pi G)$
while the second term is the well-known scalar tensor
interaction term $\phi^2 R$ with dimensionless coefficient $\xi_2=\xi$.
The running couplings are given by
\begin{equation}
\tilde\xi_{2n}=
{1\over n!}{\partial^n F\over\partial (\phi^2)^n}\Big|_{\phi=0}.
\end{equation}
As before, we define dimensionless couplings
$\xi_{2n}=k^{2(n-1)}\tilde\xi_{2n}$.
The corresponding beta functions are given by
\begin{equation}
\label{eq:beta.xi}
\partial_t \xi_{2n}=
2(n-1)\xi_{2n}+{k^{2(n-1)}\over Vol}{1\over n!}
{\partial^{n+1}\over \partial R\partial (\phi^2)^n}\partial_t \Gamma_k \Big|_{\substack{\phi=0\\ R=0}}.
\end{equation}

We now have to insert this Ansatz into the appropriate ERGE.
The derivation of equation \eqref{eq:ERGE} in the previous section
was quite general and therefore the
ERGE for gravity coupled to a scalar field has
again the same form, except for two generalizations:
first, the field $\phi$ is to be reinterpreted as a
matrix consisting of the components of the metric and a scalar
field; second, since gravity is a gauge theory, one has to take into
account the effect of gauge fixing and ghost terms.

Here we mention some points that are necessary to understand
the results; 
we refer to \cite{Dou:1998fg} and \cite{Lauscher:2001ya} for details.
In deriving the ERGE, one encounters the \emph{quantum metric} (to be
integrated out in the funtional integral), say $\gamma_{\mu\nu}$,
which can be decomposed into the sum of an arbitrary background metric
$\bar g_{\mu\nu}$ and a quantum fluctuation $h_{\mu\nu}$.
The background metric is used in the gauge fixing terms \eqref{eq:GF.action} below
and also in the cutoff terms $\Delta S_k$, which have to be quadratic
in $h_{\mu\nu}$.
In the Legendre transformation 
one encounters also the \emph{classical metric} $g_{\mu\nu}$, 
which is the canonically
conjugate variable of the source associated to the quantum metric. 
Thus, in general, the action $\Gamma_k$ will depend both on $\bar g$ and $g$.
On the other hand the Ansatz \eqref{eq:class} only depends on one metric.
In order to derive the beta functions for the couplings in \eqref{eq:class}
we proceed as follows.
In the r.h.s. of the ERGE, one first takes the functional derivatives 
w.r.t.\ the \emph{classical field} $g$, then one sets the background
metric equal to the classical one, \emph{i.e.}\ 
$g_{\mu\nu}=\bar g_{\mu\nu}$, so that many contributions disappear.
From these equations one can read off the beta functions of the couplings.

The gauge-fixing action is chosen as
\begin{gather}
  \label{eq:GF.action}
  S_{GF}=\frac{1}{2\alpha} \int\!\mathrm{d}^4 x
  \sqrt{\bar{g}}\,\bar{g}^{\mu\nu}
  F_\mu F_\nu,\\
  \hspace{2mm} F_\mu=\bar g^{\nu\rho}\left(\bar{\nabla}_\nu g_{\rho\mu}
  -\frac{\beta+1}{4}\bar{\nabla}_\mu  g_{\nu\rho}\right)\notag,
\end{gather}
so that the corresponding ghost action will be
\begin{equation}
  \label{eq:gh.action}
  S_{gh}=\int\!\mathrm{d}^4 x \sqrt{\bar{g}}\, \bar{C}_\mu
  \left(-\bar{\nabla}^2 g^{\mu\nu}+\frac{\beta-1}{2} \nabla^\mu\nabla^\nu
    -R^{\mu\nu}\right) C_\nu.
\end{equation}
In principle, $\alpha$ and $\beta$ are running parameters in the effective action,
so one should take into account their beta functions, too.
However, as will be discussed in Section VI, there are arguments
to the effect that $\alpha=0$ at the FP. Therefore,
unless otherwise stated,
we will always work in the gauge $\alpha=0$ and $\beta=1$.

The kinetic term of the gravitons is obtained by linearizing
the action around a de Sitter metric with scalar curvature $R$ 
and a constant scalar backgound $\phi$.
Using the method of \cite{Dou:1998fg}, 
the r.h.s.\ of \eqref{eq:ERGE} can be written as a sum of
several terms, corresponding to the spin
2, 1 and 0 components of the fields, and has to be completed by adding
the ghost contributions.  

The spin-2 component of the metric has an inverse propagator
\begin{equation}
\label{eq:prop2}
{1\over2}F(\phi^2)\left(z+{2\over3}R\right)-{1\over2}V(\phi^2),
\end{equation}
where now $z=-\nabla_\mu\nabla^\mu$.
The spin-1 component of the metric has an inverse propagator
\begin{equation}
\label{eq:prop1}
{1\over\alpha}F(\phi^2)\left(z+{2\alpha-1\over 4}R\right)-V(\phi^2),
\end{equation}
where $\alpha$ is the gauge-fixing parameter.
The two spin-0 components of the metric mix with the scalar field;
the resulting inverse propagator is given by the matrix
\begin{widetext}
\begin{equation}
\label{eq:prop0}
\left(
\begin{array}{c c c}
{3\over16}F(\phi^2)\left({3-\alpha\over\alpha}z
+{\alpha-1\over\alpha}R\right)-{3\over8}V(\phi^2)&
{3\over16}{\beta-\alpha\over\alpha}F(\phi^2)\sqrt{z}\sqrt{z-{R\over3}}&
-{3\over2}F'(\phi^2)\phi\sqrt{P_k}\sqrt{P_k-{R\over3}}\\
{3\over16}{\beta-\alpha\over\alpha}F(\phi^2)\sqrt{z}\sqrt{z-{R\over3}}&
-{1\over16}{3\alpha-\beta^2\over\alpha}F(\phi^2)z+{1\over8}V(\phi^2)&
-{3\over2}F'(\phi^2)\phi\left(z-{R\over3}\right)+\phi V'\\
-{3\over2}F'(\phi^2)\sqrt{z}\sqrt{z-{R\over3}}&
-{3\over2}F'(\phi^2)\left(z-{R\over3}\right)+\phi V'&
z+2V'+4\phi^2V''-R(2F'+4\phi^2F'')
\end{array}
\right).
\end{equation}
\end{widetext}
The two factors under trace in the r.h.s.\ of \eqref{eq:ERGE}
are obtained from these expressions as follows.
The modified (cutoff) propagators are given by the inverses
of the expressions in (\ref{eq:prop2},\ref{eq:prop1},\ref{eq:prop0}),
with $z$ replaced by $P_k(z)$.
The function $R_k$ for each spin component is given by the
difference of the cutoff propagator and the original propagator.
In the case of spin 2 and spin 1, this is just the function $R_k(z)$
defined in \eqref{eq:cutoff}, whereas for the spin 0 components
it is a 3$\times$3 matrix (the difference of \eqref{eq:prop0}
with $z$ replaced by $P_k(z)$ and \eqref{eq:prop0}). 

Since in (\ref{eq:prop2},\ref{eq:prop1},\ref{eq:prop0})
the momentum variable $z$ always appears multiplied by
the function $F(\phi^2)$, also the matrices
$R_k$ appearing in \eqref{eq:ERGE} contain $F(\phi^2)$.  
When inserted in the
r.h.s.\ of \eqref{eq:beta.xi}, besides the explicit dependence of
$P_k(z)$ on $k$, one has to take into account the dependence on $k$ of
all coupling constants that are present in $F(\phi^2)$ and its
derivatives (this is related to the ``renormalization group
improvement'' that turns the one-loop RG into an exact equation).
This generates terms proportional to the beta functions in the r.h.s.
of the equations, so that the ERGE does not immediately yield expressions
for the beta functions but rather linear equations for the beta
functions\footnote{ This did not happen in the pure scalar case
  because there the propagator was fixed to be equal to one. It would
  have happened if we had written a more general action containing a
  term ${1\over2}Z(\phi^2)g^{\mu\nu}\partial_\mu\phi\partial_\nu\phi$.
  Then the expressions for all beta functions would contain on the
  r.h.s.  the beta functions of the couplings that appear in the
  function $Z(\phi^2)$.}.

The beta functions themselves are then obtained by inverting the
matrix of coefficients, and this introduces further nonlinearities into
the system. We will not write the expressions for the beta functions
themselves but only the
linear equations that determine the beta functions.
We will order the
couplings in order of decreasing mass dimension (before dividing by
powers of $k$): $\l0,\x0,\l2,\x2,\l4,\x4\ldots$.  
These are the first five equations:

\begin{widetext}
  \begin{align}
    \label{eq:beta.lambda0.grav}
    \partial_t\l0=\,& \frac{1}{32\pi^2}\left\{\Qb{\Pt (\l0(3\P+8\l2) +\P (3\P+4 \l2)\x0)}
      {\P(\P+2\l2)(\x0 \P-\l0)}
      +2\frac{\x0'}{\x0}\Qb{\R(2 \l0-5\x0\P)}{\P(\l0-\x0\P)}\right\},\\[1mm]
    \label{eq:beta.xi0.grav}
    \partial_t\x0=\,& \frac{1}{384\pi^2}
    \left\{\Qa{\Pt(-\l0(3\P+10\l2)+\P(11\P+26\l2)\x0)}{\P(\P+2\l2)(\x0\P-\l0)}\right.\notag\\
    &\qquad\qquad-\Qb{\Pt(6\l0\x0\P(20\l2\P+20\l2^2+\P^2(5-8\x2))+3\l0^2(-20\l2\P-20\l2^2+\P^2(8\x2-5)))} {\P^2(\P+2\l2)^2(\x0\P-\l0)^2}\notag\\
    &\qquad\qquad+\left.\Qb{\Pt(\x0^2\P^2(-220\l2\P-220\l2^2+\P^2(24\x2-55)))}
      {\P^2(\P+2\l2)^2(\x0\P-\l0)^2}\right\}\notag\\
    &+\frac{1}{384\pi^2}\frac{\partial_t\x0}{\x0}\left\{
      \Qa{\R(5\l0+3\x0\P)}{\P(\x0\P-\l0)}
      +5\Qb{\R(-3\l0^2+6\l0\x0\P+5\x0^2\P^2)}{\P^2(\x0\P-\l0)^2}\right\},\\[1mm]
    \label{eq:beta.lambda2.grav}
    \partial_t\l2=\,& -\frac{3}{16\pi^2}
    \Qb{\Pt(2\l0^2\l4 +\x0 \P^2(2\l4\x0-\l2(1+2\x2)^2)+\l0((1+2\x2)(-2\l2^2+4\l2\x2\P+\x2\P^2)-4\l4\x0\P))}
    {(\P+2\l2)^2(\x0\P-\l0)^2}\notag\\
    &-\frac{\partial_t\x0}{16\pi^2\x0^2}\Qb{\R(-\l2(-4\l0^2\x2+8\l0\x0\x2\P+\x0^2\P^2(3+2\x2))
      +\x2\P(2\l0^2-4\l0\x0\P+(5+6\x2)\x0^2\P^2))}
    {\P(\P+2\l2)(\x0\P-\l0)^2}\notag\\
    &+\frac{\partial_t\x2}{16\pi^2 \x0}\Qb{\R(-2\l0 (\P+2\l2)+\x0\P(5\P-2\l2+12\x2\P))}
    {\P(\P+2\l2)(\x0\P-\l0)},\\[5mm]
    \label{eq:beta.xi2.grav}
    \partial_t\x2=\,&\frac{1}{48\pi^2}\Qabis{\frac{\Pt}{(\P+2\l2)^2(\x0\P-\l0)^2}}\cdot\right.\notag\\
  &\qquad\qquad\cdot\left\{3\l0^2\l4-\l0(4\l2\P(1-3\x2)\x2+\l2^2(3+16\x2)+\P(6\l4\l0+\P(1-3\x2)\x2))\right.\notag\\
  &\qquad\qquad\quad+\left.\x0(10\l2^2\P+10\l2^3+3\l4\x0\P^2+\l2\P^2(1-6\x2-6\x2^2))\right\}\bigg]\notag\\
  &-\frac{1}{48\pi^2}\Qbbis{\frac{\Pt}{\P^2(\P+2\l2)^3(\x0\P-\l0)^3}}\cdot\right.\notag\\
&\qquad\qquad\qquad\cdot\left\{-18\l0^3\P^2(-4\l4\x2+(\P+2\l2)\x4)-6\l0^2(3\l2^3\P+2\l2^4+\l2^2\P^2(1+4\x2+10\x2^2))\right.\notag\\
&\qquad\qquad\qquad\quad+6\l0^2(9\x0\P^3(-4\l4\x2+\x4\P)+\l2\P^3(4\x2+19\x2^2+24\x2^3+18\x0\x4))\notag\\
&\qquad\qquad\qquad\quad+\l0\x0\P(36\l2^4+8\l2^3\P(6+7\x2)-12\l2\P^3(\x2+18\x2^2+18\x2^3+9\x0\x4))\notag\\
&\qquad\qquad\qquad\quad+\l0\x0\P^3(3\l2^2(5+36\x2+20\x2^2)+\P(216\x0\l4\x2+\P(10\x2+21\x2^2+36\x2^3-54\x0\x4)))\notag\\
&\qquad\qquad\qquad\quad-\x0^2\P^2(104\l2^4+3\l2^2\P^2(23-12\x2)-6\l2^3\P(-25+4\x2))\notag\\
&\qquad\qquad\qquad\quad+2\x0^2\l2\P^5(-5+24\x2+51\x2^2+36\x2^3+18\x0\x4)\notag\\
&\qquad\qquad\qquad\quad+3\x0^2\P^5(-24\l4\x0\x2-7\x2^2\P-12\x2^3\P+6\x0\x4\P)\notag\bigg]\\
&+\frac{\partial_t\x0}{384\pi^2\x0^2}\Qabis{\frac{\R}{\P(\P+2\l2)(\x0\P-\l0)^2}}\cdot\right.\notag\\
&\qquad\qquad\qquad\cdot\left\{40\l2^2\x0^2\P+\x2\P(5\l0^2-10\l0\x0\P+3\x0^2\P^2(-1+8\x2))\right.\notag\\
&\qquad\qquad\qquad\quad-\left.2\l2(-5\l0^2\x2+10\l0\x0\x2\P+\x0^2\P^2(-4+27\x2))\right\}\bigg]\notag\\
&-\frac{\partial_t\x0}{384\pi^2\x0^2}\Qbbis{\frac{\R}{\P^2(\P+2\l2)^2(\x0\P-\l0)^3}}\cdot\right.\notag\\
&\qquad\qquad\qquad\cdot\left\{15\l0^3(\P+2\l2)^2\x2-3\l0^2\x0(\P+2\l2)(8\l2^2+15\x2\P^2+22\l2\x2\P)\right.\notag\\
&\qquad\qquad\qquad\quad+\x0^3\P^2(-416\l2^3+4\l2^2\P(-92+61\x2))\notag\\
&\qquad\qquad\qquad\quad+\x0^3\P^4(\x2\P(25-168\x2-288\x2^2)+4\l2(-20+79\x2+60\x2^2))\notag\\
&\qquad\qquad\qquad\quad+\l0\x0^2\P(144\l2^3-20\l2\x2\P^2(-5+12\x2))\notag\\
&\qquad\qquad\qquad\quad+\left.\l0\x0^2\P^2(4\l2^2(18+37\x2)+\x2\P^2(85+168\x2+288\x2^2))\right\}\bigg]\notag\\
&-\frac{\partial_t\x2}{384\pi^2\x0}\Qabis{\frac{\R}{\P(\P+2\l2)(\x0\P-\l0)}}\cdot
\left\{-5\l0(\P+2\l2)+3\x0\P(-\P-18\l2+16\x2\P)\right\}\right]\notag\\
&-\frac{\partial_t\x2}{384\pi^2\x0}\Qbbis{\frac{\R}{\P^2(\P+2\l2)^2(\x0\P-\l0)^2}}\cdot\right.\notag\\
&\qquad\qquad\qquad\cdot\left\{15\l0^2(\P+2\l2)^2-6\l0\x0\P(12\l2^2+8\l2\P(2-5\x2)+\P^2(5+28\x2+96\x2^2))\right.\notag\\
&\qquad\qquad\qquad\quad+\left.\x0^2\P^2(-148\l2^2-4\l2\P(31+60\x2)+\P^2(-25+168\x2+576\x2^2))\right\}\bigg],\\
\label{eq:beta.lambda4.grav}
\partial_t\l4=\,&\frac{3}{16\pi^2}\Qbbis{\frac{\Pt}{(\P+2\l2)^3(\x0\P-\l0)^3}}\cdot\right.\notag\\
&\qquad\qquad\cdot\left\{\l0^3(-24\l4^2+5(\P+2\l2)\l6)+\l0^2(-24\l2^3\x4+4\l2^2(8\l4(1+2\x2)+\P(-1+16\x2)\x4))\right.\notag\\
&\qquad\qquad\quad+\l0^2\P(72\l4^2\x0+\P(-15\l6\x0+\x4\P(1+8\x2)))\notag\\
  &\qquad\qquad\quad+2\l0^2\l2\P((-4\l4(1+8\x2+12\x2^2)+3(-5\l6\x0+\P(1+8\x2)\x4)))\notag\\
  &\qquad\qquad\quad+\x0(40\l2^5-40\P\l2^4(-1+2\x2)-24\P\l2^3(-\l4\x0+5\x2\P))\notag\\
  &\qquad\qquad\quad+\x0^2\P^3(24\l4^2\x0-5\l6\x0\P+\l4\P(1+8\x2+24\x2^2))\notag\\
  &\qquad\qquad\quad+\x0\l2\P^3(-10\l6\x0^2-2\l4\x0(5+40\x2+24\x2^2)-\P(6\x2^2+52\x2^3+48\x2^4-8\x0\x4+\x2(1-16\x0\x4)))\notag\\
  &\qquad\qquad\quad+\l0(20\l2^4(1+4\x2)-2\l2^3(-12\l4\x0+\P(1+24\x2+124\x2^2+12\x0\x4)))+\notag\\
  &\qquad\qquad\quad+2\l0\l2^2\P(2\l4\x0(17+16\x2)+3\P(\x2+4\x2^2+44\x2^3+2\x0\x4+16\x0\x2\x4))\notag\\
  &\qquad\qquad\quad+\l0\l2\P^2(-30\l6\x0^2-18\l4\x0(1+8\x2+8\x2^2)+\P(28\x2^3-96\x2^4+14\x0\x4+\x2(1+64\x0\x4)))\notag\\
  &\qquad\qquad\quad\left.+\l0\P^2(72\l4^2\x0^2+\P\l4\x0(1+8\x2+24\x2^2)
    +\P(-15\l6\x0^2-\P(\x2+4\x2^3+24\x2^4-\x0\x4-8\x0\x2\x4)))\right\}\!\bigg]\notag\\
  &+\frac{\partial_t\x0}{16\pi^2\x0^3}\Qbbis{\frac{\R}{\P(\P+2\l2)^2(\x0\P-\l0)^3}}\cdot\right.\notag\\
&\qquad\qquad\qquad\cdot\left\{(60\l2^4\x0^3\P-240\l2^3\x0^3\x2\P^2+\l2^2(-8\l0^3\x2^2+4\x0^4\P^2(9\l4+7\x4\P)-24\l0\x0^2\P(\x2^2\P+\l0\x4)))\right.\notag\\
&\qquad\qquad\qquad+\l2^2(8\l0^2\x0(3\x2^2\P+\l0\x4)+\x0^3\P(\P^2(3+12\x2+380\x2^2)-12\l0(3\l4+\x4\P)))\notag\\
&\qquad\qquad\qquad+\P^2(-2\l0^3+6\l0^2\x0\P)(\x2^2-\x0\x4)\notag\\
&\qquad\qquad\qquad+\P^3(\x0^3\P(3\l4\x0(1+8\x2+24\x2^2)+5\x2^2+12\x2^3+72\x2^4-5\x0\x4-24\x0\x2\x4))\notag\\
&\qquad\qquad\qquad+3\l0\x0^2\P^3(-\l4\x0(1+8\x2+24\x2^2)+\P(-2\x2^2+3\x0\x4+8\x0\x2\x4))\notag\\
&\qquad\qquad\qquad-2\l2\P((4\l0^3-12\l0^2\x0\P)(\x2^2-\x0\x4)-6\l0\x0^2\P(\l4\x0(1+8\x2)+\P(-2\x2^2+\x0\x4+4\x0\x2\x4)))\notag\\
&\qquad\qquad\qquad\left.-2\l2\x0^3\P^2(6\l4\x0(1+8\x2)+\P(8\x2^2+132\x2^3-2\x0\x4+\x2(3+24\x0\x4)))\right\}\!\bigg]\notag\\
&+\frac{\partial_t\x2}{16\pi^2\x0^2}\Qbbis{\frac{\R}{\P(\P+2\l2)^2(\x0\P-\l0)^2}}\cdot\right.\notag\\
&\qquad\qquad\qquad\cdot\left\{120 \l2^3\x0^2\P-2\l2^2(4\l0^2\x2-8\l0\x0\x2\P+\x0^2\P^2(3+190\x2))\right.\notag\\
&\qquad\qquad\qquad+\l2\P(-8\l0^2\x2+16\l0\x0\x2\P+48\x0^3\P(2\l4+\x4\P))\notag\\
&\qquad\qquad\qquad+\x0^2(\P^2(3+16\x2+396\x2^2)-48\l0(2\l4+\x4\P)))\notag\\
&\qquad\qquad\qquad+\P^2(-2\l0^2\x2-\x0^2\P(24\l4\x0(1+6\x2)+\P(\x2+18\x2^2+144\x2^3-24\x0\x4)))\notag\\
&\qquad\qquad\qquad+\left.4\l0\x0\P^2(6\l4\x0(1+6\x2)+\P(\x2-6\x0\x4))\right\}
\!\bigg]\notag\\
&+\frac{\partial_t\x4}{16\pi^2\x0}\Qbbis{\frac{\R}{(\P+2\l2)^2(\x0\P-\l0)^2}}\cdot
\left\{-2\l0(\P+2\l2)+\x0\P(5\P-14\l2+24\x2\P)\right\}\right].
\end{align}
\end{widetext}

This system of linear equations can be solved for
$\beta^\lambda_{2i}=\partial_t\lambda_{2i}$
and
$\beta^\xi_{2i}=\partial_t\xi_{2i}$.
Note that we have not made any truncation on the functions $V$ and $F$:
no couplings have been assumed to be zero.

We do not exhibit the equations for the higher couplings.
By means of algebraic manipulation software we have calculated
the beta functions up to $\l8$ and $\x6$. 
The general pattern is however already clear from the equations shown here.
The functions $Q_1$ and $Q_2$ always contain denominators involving only the
couplings $\l0$, $\l2$ and $\x0$.
Aside from these couplings appearing inside $Q_1$ and $Q_2$,
the equation for $\beta^\lambda_{2i}$ is a polynomial in the couplings
$\l0, \l2,\ldots,\l{2i+2}$ and $\x0,\x2,\ldots,\x{2i}$,
with the functions $Q_1$ and $Q_2$ as coefficients,
while the equation for $\beta^\xi_{2i}$ is a polynomial in the couplings
$\l0, \l2,\ldots,\l{2i+2}$ and $\x0,\x2,\ldots,\x{2i+2}$,
with the functions $Q_1$ and $Q_2$ as coefficients.
When these equations are solved, the beta function
$\beta^\lambda_{2i}$
is written as a rational function of
$\l0, \l2,\ldots,\l{2i+2}$ and $\x0,\x2,\ldots,\x{2i}$,
while the beta function
$\beta^\xi_{2i}$
is written as a rational function of
$\l0, \l2,\ldots,\l{2i+2}$ and $\x0,\x2,\ldots,\x{2i+2}$.

It is clear that the system 
$\beta^\lambda_{2i}=0$, $\beta^\xi_{2i}=0$
admits a FP for which all couplings $\lambda_{2i}$ and
$\xi_{2i}$ vanish for $i>0$, while for $i=0$
$\lambda_0=2\kappa\Lambda$ and $\xi_0=\kappa$ have the same
values that they would have in the presence of a single free scalar
field, as discussed in \cite{Percacci:2002ie} (these values are
numerically very close to those of pure gravity, discussed in
\cite{Lauscher:2001ya}), namely, for $a=2$,
\begin{equation}
  \label{eq:FP.values}
    \begin{aligned}
      \l0{}_*&= 0.0080022,\\
      \x0{}_*&= 0.023500.
    \end{aligned}
\end{equation}
To compare with the results of \cite{Lauscher:2001ya}, 
we define the dimensionless variables
$\lambda=\Lambda/k^2=\l{0}/2\x{0}$
and $g=G k^2=1/16\pi\x{0}$. At the GMFP
\begin{equation}
  \label{eq:FP.values.old.notation}
    \begin{aligned}
      \lambda_*&=\frac{\l{0*}}{2\x{0*}}= 0.1703,\\
      g_*&= \frac{1}{16 \pi\x{0*}}= 0.8466.
    \end{aligned}
\end{equation}
These values differ from those in eq.\ (5.25) of \cite{Lauscher:2001ya}
on two accounts: they are calculated for different values of
the cutoff parameter $a$, and here the FP is shifted due to the presence
of the scalar field.
When these factors are taken into account there is perfect agreement.
(See figure (5.b) of \cite{Lauscher:2001ya} for the dependence of
results on $s=2a$ and compare with figure \ref{fig:cutoff_indep}.)

This FP can be viewed alternatively as the FP of pure gravity,
slightly shifted due to the presence of a free, massless, minimally
coupled scalar, or as the Gaussian FP of the pure scalar theory,
generalized to include gravitational interactions.  It is remarkable
that matter remains ``non-self-interacting'' at this FP, and that the
only nonzero couplings are those that affect only the gravitational
degrees of freedom.  
(This goes some way towards justifying the assumption in 
\cite{Percacci:2002ie} that matter fields are non-self-interacting.)
For want of a better terminology, we shall refer
to this FP as the Gaussian-Matter FP, of GMFP for short.

The issue arises whether the coupled system of equations
admits other nontrivial FP's.
The complexity of the equations has prevented us from
deriving definite results on this issue.
We have looked for other FP's using numerical methods
in a five-parameter truncation of the theory containing
the couplings $\lambda_{2n}$ for $n=0,1,2$ and the couplings
$\xi_{2n}$ for $n=0,1$.
Our method consists in considering a grid in the space of all
parameters, evaluating the beta functions by numerical integration at
a point and then at all neighboring points.  If all beta functions
change sign simultaneously when going from a point to a neighbor,
then generically there will be a FP somewhere 
near the link between the two points. The region is then examined with
a finer grid until the position of the FP is located with sufficient
accuracy.
We have started off with the $2\times 2$ grid given by $\l0$ and
$\x0$, confirming the results of \cite{Lauscher:2001ya}; we
have then added one by one the other variables involved in the
five-parameter truncation, getting increasingly complicated systems
of equations. 


Because of the complexity of the beta functions, the numerical
evaluation takes considerable time. The largest range we have explored
is a $5\times 5$ grid with the dimensionless cosmological constant
$\lambda$ ranging from 0.010 to 0.045 in steps
of 0.005; the dimensionless Newton constant $g$
ranging from 0.01 to 0.06 in steps of 0.01; the dimensionless scalar
mass $2\l2$ ranging from -1 to 1 in steps of 0.2; the
dimensionless quartic scalar coupling $\l4$ ranging from -15
to 5 in steps of 1; the dimensionless scalar-tensor coupling
$\xi=\x2$ ranging from -5 to 5 in steps of 1.  This makes a
lattice with more than 120,000 points. Many other attempts have been
tried with finer lattices and/or fewer parameters. 
We did find some nontrivial solutions when considering less than 5
parameters, but none of them survived the addition of an extra
parameter, so that we had to conclude that they were all spurious FPs 
due to the truncation. 
As a further check we have also resorted to series expansions around 
the GMFP or one of the spurious FP mentioned above. 
All the results obtained in this way are perfectly consistent
with the other calculations.
The outcome of all these efforts is that no FP other than the GMFP was found.

This is of course not a proof that an FP does not exist in this range
of couplings. For example, a beta function may change sign twice on
a link, once from positive to negative and once from negative to
positive, and if the distance between the zeroes is smaller than the
size of the step, it may well escape detection by our methods.
Nevertheless, after this numerical work, we consider it quite unlikely
that another nontrivial FP exists in the range of values for the
couplings that we have considered.

This result is corroborated by the following observation.
If one does not
truncate the functions $V$ and $F$ to polynomials, as in the pure
scalar case the structure of the beta functions seems to allow
for a recursive solution depending on two free parameters.  If we fix
arbitrary values for $\lambda_0$ and $\xi_0$, from equations
(\ref{eq:beta.lambda0.grav},\ref{eq:beta.xi0.grav}) one derives
$\lambda_2$ and $\xi_2$; substituting them into equations
(\ref{eq:beta.lambda2.grav},\ref{eq:beta.xi2.grav}), one can solve for
$\lambda_4$ and $\xi_4$ and so on.  This will determine the functions
$V=V_*$ and $F=F_*$ up to two arbitrary parameters. It will be
interesting to analyze this in detail and to see whether the resulting
functions $V_*$ and $F_*$ are regular or still present the problems
discussed in \cite{Morris:1996nx}.
In any case, it seems highly unlikely that the solutions will
be polynomial.
This point of view also sheds a different light on the FP
found in \cite{Souma:1999at,Lauscher:2001ya}. The values of
$\lambda_{0*}$ and $\xi_{0*}$ at the GMFP are the only ones for
which $\l{2*}=0$ and $\x{2*}=0$, and as a consequence all the higher
couplings turn out to be zero, in accordance with the truncation
made there.  We shall not pursue this issue anymore here.  In the
rest of this paper we shall restrict our attention to the GMFP, which
is a special member of this family of solutions, and is definitely a
physically acceptable solution.

\section{Linearized flow around the GMFP}
Having established the existence of the GMFP in the truncation
defined by the action \eqref{eq:class}, we have to study its
properties, in particular to find the dimension of the critical surface.

We begin by calculating the matrix $M_{ij}$.  Again, we order the
couplings in order of decreasing mass dimension: 
$\l0,\x0,\l2,\x2,\l4,\x4,\l6,\ldots$.  
As in the pure scalar theory, due to the functional dependences
of $\beta^\lambda_{2i}$ and $\beta^\xi_{2i}$ on the couplings,
an infinite triangle above the diagonal is zero.
Furthermore, due to the fact that only
the ``purely gravitational'' couplings $\l0$ and $\x0$ are nonzero at
the GMFP, an infinite triangle below the diagonal is zero.
The structure of the matrix $M$ is therefore remarkably simple:
\begin{equation}
\label{eq:stab.matrix.grav}
\left(
\begin{array}{c c c c c}
M_{00}&
M_{02}&
0&
0&
\cdots\\
0&
M_{22}&
M_{24}&
0&
\cdots\\
0&
0&
M_{44}&
M_{46}&
\cdots\\
0&
0&
0&
M_{66}&
\cdots\\
\cdots&
\cdots&
\cdots&
\cdots&
\cdots
\end{array}
\right)
\end{equation}
where each one of the nonzero entries is a $2\times 2$ matrix
of the form

\begin{equation}
M_{ij}=
\left(
\begin{array}{c c}
{\partial\beta^\lambda_{(2i)}\over\partial\lambda_{(2j)}}&
{\partial\beta^\lambda_{(2i)}\over\partial\xi_{(2j)}}\\
{\partial\beta^\xi_{(2i)}\over\partial\lambda_{(2j)}}&
{\partial\beta^\xi_{(2i)}\over\partial\xi_{(2j)}}
\end{array}
\right).
\end{equation}

For the calculation of the dimension of the critical surface
we need to count the number of negative eigenvalues of the matrix $M$.
The eigenvalue problem for the matrix $M$ could be turned into
recursion relations for $\l{2i}$ and $\x{2i}$,
as for the pure scalar theory.
However, if we restrict ourselves to solutions where $V$ and $F$
are polynomials,
given the almost-block-diagonal structure of $M$, 
the eigenvalues of $M$ are just the eigenvalues 
of the diagonal blocks $M_{ii}$.
Explicitly, the diagonal blocks have the following form:
\begin{equation}
M_{ii}=
\left(
\begin{array}{c c}
2(i-2)&
0\\
0&
2(i-1)\\
\end{array}\right)
+
\left(
\begin{array}{c c}
\delta M_{\lambda\lambda}&
\delta M_{\lambda\xi}\\
\delta M_{\xi\lambda}&
\delta M_{\xi\xi}\\
\end{array}\right),
\end{equation}
where the first term contains the canonical dimensions of the
couplings and the second term, which contains the quantum corrections,
has the following form:
\begin{widetext}
\begin{multline}
 \delta M_{\lambda\lambda}= \frac{\x0}{16\pi^2}\left\{
3\Qb{\Pt}{(\l0-\P\x0)^2}-\frac{1}{\Delta^2}\cdot\right.\\
  \hspace{3mm}\cdot\left[{}-3\Qb{\R}{(\l0-\x0\P)^2}\cdot
    \left(\Qa{\Pt(-3\l0+11\x0\P)}{\P(-\l0+\x0\P)}
      +5\Qb{\Pt(3\l0^2-6\l0\x0\P+11\x0^2\P^2)}{\P^2(\l0-\x0\P)^2}\right)\cdot \Delta\right.\\
  \hspace{8mm}-8\Qb{\R(-2\l0+5\x0\P)}{\P(-\l0+\x0\P)}\cdot
  \left(\Qa{\Pt}{(\l0-\x0\P)^2}+10\x0\Qb{\Pt}{(\x0\P-\l0)^3}\right)\cdot\Delta\\
  \hspace{8mm}-8\Qb{\R(-2\l0+5\x0\P)}{\P(-\l0+\x0\P)}
\cdot\left(\Qa{\R}{(\l0-\x0\P)^2}+10\x0\Qb{\R}{(\x0\P-\l0)^3}\right)\\
    \left.\left.\cdot\left(\Qa{\Pt(-3\l0+11\x0\P)}{\P(-\l0+\x0\P)}
    +5\Qb{\Pt(3\l0^2-6\l0\x0\P+11\x0^2\P^2)}{\P^2(\l0-\x0\P)^2}\right)\right]\right\},
\end{multline}
\begin{multline}
 \delta M_{\lambda\xi}=-\frac{3\l0}{16\pi^2}\Qb{\Pt}{(\l0-\x0\P)^2}\\
  \hspace{8mm}{}-\frac{1}{8\pi^2\Delta}
  \mathrm{Q_2}\!\left[\R\left(\frac{2}{\x0\P}+\frac{3}{\x0\P-\l0}\right)\right]\left\{\Qa{\Pt(-3\l0+11\x0\P)}{\P(-\l0+\x0\P)}
    +5\Qb{\Pt(3\l0^2-6\l0\x0\P+11\x0^2\P^2)}{\P^2(\l0-\x0\P)^2}\right\}\\
  -\frac{1}{32\pi^2\Delta^2}\left\{6\l0\x0\cdot\Delta\cdot\Qb{\R}{(\l0-\x0\P)^2}\cdot\left(\Qa{\Pt(-3\l0+11\x0\P)}{\P(-\l0+\x0\P)}
      +5\Qb{\Pt(3\l0^2-6\l0\x0\P+11\x0^2\P^2)}{\P^2(\l0-\x0\P)^2}\right)\right.\\
  \hspace{8mm}{}-2\Qb{\R(-2\l0+5\x0\P)}{\P(-\l0+\x0\P)}\cdot\left(\Qa{\Pt(-3\l0+11\x0\P)}{\P(-\l0+\x0\P)}
    +5\Qb{\Pt(3\l0^2-6\l0\x0\P+11\x0^2\P^2)}{\P^2(\l0-\x0\P)^2}\right)\cdot\\
  \hspace{25mm}{}\cdot\left(\Qa{\Pt(5\l0^2-10\l0\x0\P-3\x0\P^2)}{\P(\l0-\x0\P)^2}
    -5\Qb{\R(3\l0^3-9\l0^2\x0\P+17\l0\x0^2\P+5\x0^3\P^3)}{\P^2(-\l0+\x0\P)^3}\right)\\
  \hspace{8mm}{}-2\cdot\Delta\cdot\Qb{\R(-2\l0+5\x0\P)}{\P(-\l0+\x0\P)}\cdot\\
   \hspace{11mm}{} \left.\left(\Qa{\Pt(3\l0^2-22\x0\l0\P+11\x0^2\P^2)}{\P(\l0-\x0\P)^2}
    +5\Qb{\Pt(-3\l0^3+9\l0^2\x0\P-33\l0\x0^2\P^2+11\x0^3\P^3)}{\P^2(-\l0+\x0\P)^3}\right)\right\},
\end{multline}
\begin{multline}
 \delta M_{\xi\lambda} = -2\frac{\x0}{\Delta^2}\cdot
  \hspace{8mm}\left\{-4\x0\left(\Qa{\Pt}{(\l0-\x0\P)^2}+10\x0\Qb{\Pt}{(-\l0+\x0\P)^3}\right)\cdot\Delta\right.\\
  \hspace{10mm}-4\x0\left(\Qa{\Pt(-3\l0+11\x0\P)}{\P(-\l0+\x0\P)}
    +5\Qb{\Pt(3\l0^2-6\l0\x0\P+11\x0^2\P^2)}{\P^2(\l0-\x0\P)^2}\right)\cdot\\
  \hspace{8mm}\left.\left(\Qa{\R}{(\l0-\x0\P)^2}+10\Qb{\R}{(-\l0+\x0\P)^3}\right)\right\},\hspace{2cm}
\end{multline}
\begin{multline}
  \delta M_{\xi\xi}=-\frac{1}{\Delta^2}\cdot
  \hspace{3mm}\cdot\left\{-\left(\Qa{\Pt(-3\l0+11\x0\P)}{\P(-\l0+\x0\P)}
      +5\Qb{\Pt(3\l0^2-6\l0\x0\P+11\x0^2\P^2)}{\P^2(\l0-\x0\P)^2}\right)\cdot\right.\\
  \hspace{15mm}\cdot\left(\Qa{\R(5\l0-10\l0\x0\P-3\x0^2\P^2)}{\P(\l0-\x0\P)^2}
    -5\Qb{\R(3\l0^3-9\l0^2\x0\P+17\l0\x0^2\P^2+5\x0^3\P^3)}{\P^2(-\l0+\x0\P)^3}\right)\\
  \hspace{8mm} +\left(\Qa{\Pt(-3\l0+11\x0\P)}{\P(-\l0+\x0\P)}
    +5\Qb{\Pt(3\l0^2-6\l0\x0\P+11\x0^2\P^2)}{\P^2(\l0-\x0\P)^2}\right)\cdot\Delta\\
  \hspace{6mm} -\left(\Qa{\Pt(3\l0^2-22\l0\x0\P+11\x0^2\P^2)}{\P(\l0-\x0\P)^2}
  \left.-5\Qb{\Pt(-3\l0^3+9\l0^2\x0\P-33\l0\x0^2\P^2+11\x0^3\P^3)}{\P^2(-\l0+\x0\P)^3}\right)\cdot \Delta\right\},
\end{multline}
where
\begin{equation}
\Delta=\Qa{\R(5\l0+3\x0\P)}{\P(\l0-\x0\P)}
+5\Qb{\R(3\l0^2-6\l0\x0\P-5\x0^2\P^2)}{\P^2(\l0-\x0\P)^2}+384\pi^2\x0\ .
\end{equation}
\end{widetext}

The most remarkable property of these quantum corrections is
that they are independent of $i$,
so that the eigenvalues of $M_{2i2i}$ simply
grow by 2 whenever $i$ is increased by 1.
For example, choosing the cutoff with $a=2$, we have the
following numerical results:

\begin{equation}
M_{00}=
\left(
\begin{array}{c c}
1.1257&
-2.5192\\
8.1295&
-5.3604\\
\end{array}
\right),
\end{equation}
which has eigenvalues $-2.1173\pm 3.1563 i$;
\begin{equation}
M_{22}=
\left(
\begin{array}{c c}
3.1257&
-2.5192\\
8.1295&
-3.3604\\
\end{array}
\right),
\end{equation}
with eigenvalues $-0.1173\pm 3.1563 i$;
\begin{equation}
M_{44}=
\left(
\begin{array}{c c}
5.1257&
-2.5192\\
8.1295&
-1.3604\\
\end{array}
\right),
\end{equation}
with eigenvalues $1.8826\pm 3.1563 i$;
\begin{equation}
M_{66}=
\left(
\begin{array}{c c}
7.1257&
-2.5192\\
8.1295&
0.6396\\
\end{array}
\right),
\end{equation}
with eigenvalues $3.8826\pm 3.1563 i$,
and so on.

The off-diagonal blocks $M_{i\,i+1}$ in \eqref{eq:stab.matrix.grav} do
not affect the eigenvalues but determine the mixing
between the couplings. Numerically, we have
\begin{equation}
M_{02}=
\left(
\begin{array}{c c}
-0.005036&
-0.002264\\
0.002736&
-0.007585\\
\end{array}
\right),
\end{equation}
\begin{equation}
M_{24}=
\left(
\begin{array}{c c}
-0.03021&
-0.01359\\
0.01642&
-0.04551\\
\end{array}
\right),
\end{equation}
\begin{equation}
M_{46}=
\left(
\begin{array}{c c}
-0.07554&
-0.03340\\
0.04104&
-0.1138\\
\end{array}
\right),
\end{equation}
and so on.

The first two (complex conjugate) eigenvectors have components
\begin{equation}
\left(
\begin{array}{c}
-0.3486\pm 0.3392 i\\
0.8737\\
0\\
0\\
0\\
0\\
0\\
0\\
\ldots\\
\end{array}
\right).
\end{equation}
They are a mixing of $\lambda_0$ and $\xi_0$;
the corresponding (complex conjugate) eigenvalues have 
negative real part $-2.1173$ and therefore these are 
relevant couplings.
 
The second and third eigenvectors have components
\begin{equation}
\left(
\begin{array}{c}
(-16.59\mp 5.343 i)\times 10^{-4}\\
(-2.970\pm 1.136 i)\times 10^{-3}\\
-0.3485\pm 0.3392 i\\
0.8736\\
0\\
0\\
0\\
0\\
\ldots\\
\end{array}
\right).
\end{equation}
They are essentially a mixing of $\lambda_2$ and $\xi_2$, with small
contributions from $\l0$ and $\x0$;
the corresponding (complex conjugate) eigenvalues have 
negative real part $-0.1173$ and therefore also these couplings are relevant.
Since they lie very close to the plane spanned by the relevant
coupling $\l2$ (canonical dimension 2) 
and the marginal coupling $\x2$ (canonical dimension 0),
we can say by a slight abuse of language that the quantum corrections
change the dimension of $\l2$ and $\x2$ making them both relevant.

The fifth and sixth eigenvector have components
\begin{equation}
\left(
\begin{array}{c}
(19.34\mp 2.834 i)\times 10^{-6}\\
(2.653\mp 2.310 i)\times 10^{-5}\\
(-9.951\mp 3.203 i)\times 10^{-3}\\
(-17.81\pm 6.820 i)\times 10^{-3}\\
-0.3485\pm 0.3392 i\\
0.8727\\
0\\
0\\
\ldots\\
\end{array}
\right).
\end{equation}
They are essentially a mixing of $\lambda_4$ and $\xi_4$, with small
contributions from $\l0$, $\x0$, $\l2$ and $\x2$;
the corresponding (complex conjugate) eigenvalues have 
positive real part 1.8826 and therefore these couplings are irrelevant.
Since they lie very close to the plane spanned by 
the marginal coupling $\l4$ (canonical dimension 0)
and the irrelevant coupling $\x4$ (canonical dimension -2),
we can say by a slight abuse of language that the quantum corrections
change the dimension of $\l4$ and $\x4$ making them both irrelevant.

The pattern continues.
The eigenvalues come in complex conjugate pairs,
and are formed by mixing the couplings 
$\lambda_{2i}$ and $\xi_{2i}$, 
with small contributions from the lower couplings.
The eigenvalues also occur in complex conjugate pairs,
and are equal to (minus) the canonical dimensions
of the couplings ($2(i-2)$ and $2(i-1)$ respectively)
plus a quantum correction.
The correction is positive for the couplings
of the series $\lambda_{2i}$ and negative for
those of the series $\xi_{2i}$, and the resulting dimension
is always contained between those of the two main couplings that
enter into the mix.

All eigenvalues differ from the first two by multiples of 2.
In particular, all the eigenvalues from the fifth onward
have positive real parts, so that the dimension
of the critical surface is four.  
The naive expectation based on canonical dimensions would have been five
(or three, if we don't count the two marginal couplings).
The quantum corrections modify the dimension of the two marginal couplings 
$\l4$ and $\x2$ so that $\x2$ (after mixing with $\l2$) becomes relevant
while $\l4$ (after mixing with $\x4$) becomes irrelevant.

\section{Effect of other matter fields}
In this section we assume that in addition to the graviton and the scalar field
discussed in the previous sections, there are $n_S-1$ new real scalar
fields, $n_W$ Weyl fields, $n_M$ Maxwell fields, $n_{RS}$ (Majorana)
Rarita-Schwinger fields, all minimally coupled.  
We neglect all masses and interactions of these additional matter fields.
The only interactions are the ones discussed in the previous sections.
This generalizes the results of \cite{Percacci:2002ie}
where only the couplings $\l0$ and $\x0$ were taken into account.
We also give some more details of the calculations.

In the presence of these new fields, the equations
(\ref{eq:beta.lambda0.grav},\ref{eq:beta.xi0.grav}) 
for the beta functions are modified by
the addition of the following terms:
%
\begin{align}
\delta\partial_t\lambda_0&=\frac{1}{32\pi^2} \left(n_S-2n_W+2n_M-4n_{RS}\right)\Qb{\Pt}{\P},\\
\delta\partial_t\xi_0&=\frac{1}{384\pi^2}\left\{
  \left(-2n_S+4n_W-4n_M\right)\Qa{\Pt}{\P}\right.\notag
\\&\left.+\left(-6n_W+9n_M-16n_{RS}\right)\Qb{\Pt}{\P^2}\right\}.
\end{align}
%

Since the contribution of the new fields to the effective action is
independent of $\phi$, they do not affect at all the beta functions of all
couplings $\lambda_{2i}$ and $\xi_{2i}$ for $i\geq1$.

The equations for the couplings $\lambda_{2i}$ for $i\geq 1$ and $\xi_{2i}$ 
for $i\geq 1$ are automatically satisfied at the GMFP.
Therefore the only equations that remain to solve are the ones
for $\l0$ and $\x0$. 
For the sake of comparison with \cite{Percacci:2002ie} we will use the
couplings $\lambda$ and $g$ in place of $\l0$ and $\x0$.
The system of these two equations is the same as the one discussed in
\cite{Percacci:2002ie}, and therefore the values of $\l0$ and $\x0$ 
at the GMFP coincide with the ones calculated therein.

We recall some of the calculations in \cite{Percacci:2002ie}. 
For the purpose of finding the fixed points, one can use the following trick.
We observe that at a fixed point $\partial_t\x0/\x0=-\partial_t
g/g=2$.  Therefore, the equations for the fixed points are
equivalent to another, simpler, set of equations which is obtained by
replacing $\partial_t\x0/\x0$ with 2 in the r.h.s.'s of
(\ref{eq:beta.lambda0.grav},\ref{eq:beta.xi0.grav}).  
Then, the equation $\beta^\lambda=0$ can be replaced by
\begin{equation}
g \cdot c (\lambdah)- 2 \lambdah=0,
\label{eq:linear.g}
\end{equation}
where $c(\lambda)$ is obtained by formally replacing
$G$ with 1 and $\partial_t G$ with $-2$ in the expression for
$\partial_t\Lambda/k^4$.
When $c(\lambdah)\neq 0$, we can solve \eqref{eq:linear.g} for $\gh$
and substitute the result into $\beta^g=0$. We shall denote
\begin{equation}\label{eq:notation}
h(\lambdah)=\beta^{\gh}\left(\lambdah,\frac{2\lambdah}{c(\lambdah)}\right),
\end{equation}
so that the zeroes of $h$ correspond to the FP's.

\begin{figure}[b]\vspace*{3mm}
  \centering{\resizebox{\figurewidth}{!}
    {\includegraphics{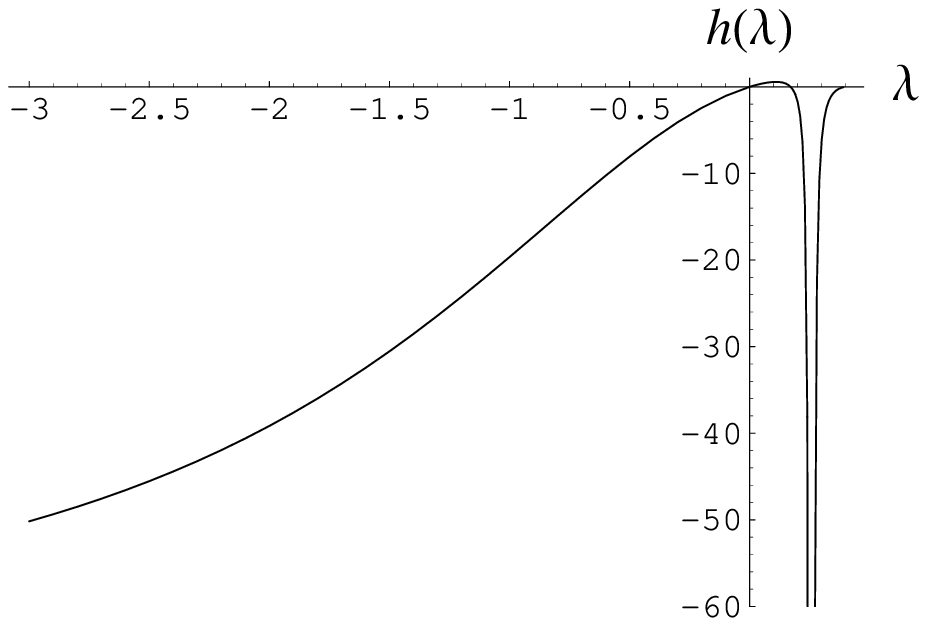}}\\[2mm]
\fbox{\resizebox{.5\figurewidth}{!}
    {\includegraphics{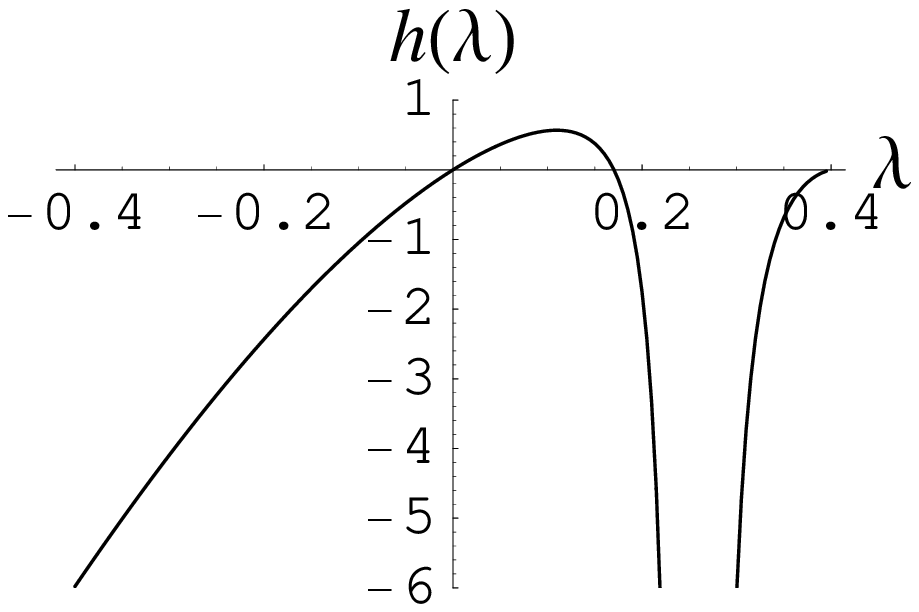}}}
    }
\caption{\label{fig:I.eps} Plot of the function $h$ for
$n_S=0, n_W=0, n_M=0$ (Region I); the asymptotic value of $h$ 
for $t\to -\infty$ is -65.3.}%
\end{figure}

The general behaviour of the function $h$ is controlled by the 
values of two parameters $\Delta'$ and $\tau$, which in turn depend on
the type and number of matter fields.
The parameter $\Delta'$ is equal to $\Delta+\sigma$, where
$\Delta=n_b-n_f$ is the difference of the total numbers of
bosonic and fermionic degrees of freedom
($n_f=2n_W+4n_{RS}$ and $n_b=n_S+2n_M+2$) and
$\sigma\equiv\!20\,\Qb{\R}{\P}/\Qb{\Pt}{\P}$ is approximately
equal to $3.64$ (for $a=2$).

The value of the cosmological constant at the FP, $\lambda_*$,
is zero on the hyperplane $\Delta'=0$.
To see this, note that
when $c(\lambdah)=0$, equation \eqref{eq:linear.g} implies
$\lambdah$=0. Therefore, if $c(0)\neq 0$ the only solution with
$\lambdah_*=0$ is the Gaussian FP, but if $c(0)=0$ we can have a 
GMFP with $\lambdah_*=0$. Explicitly,
\begin{equation}
c(0)=\frac1{4\pi k^4}\!\left[\left(n_b-n_f\right)\!
\Qb{\Pt}{\P}\!+\!20\,\Qb{\R}{\P}\right],
\end{equation}
so that the condition for
the existence of a non-Gaussian FP with zero cosmological 
constant is precisely 
\begin{equation}
\label{eq:bound}
\Delta'=0\ .
\end{equation}
Due to the irrationality of $\sigma$, there is in general no combination of
matter fields that satisfies this condition; however, the hyperplane
defined by \eqref{eq:bound} has an important physical significance: it
separates the regions with positive and negative $\lambdah_*$, as
shall become clear below.

The function $h$ tends to zero when 
$\lambda \to \mathrm{min}_{z\in[0,\infty]}(\P)/2$.
However, this point does not correspond to an FP:
For this value of $\lambda$ the denominators in the functions $Q_1$ and $Q_2$
appearing in the beta functions vanish and the beta functions
themselves blow up.
Moreover, it was shown in \cite{Reuter:1998cp} that the 
Ward identities break down near this point.  
Consequently, only values of $\lambda$ strictly less than
$\mathrm{min}_{z\in[0,\infty]}(\P)/2$ will be considered.

The function $h(\lambda)$ always has a zero at the origin,
corresponding to the Gaussian FP.
The derivative of $h(\lambda)$ at the origin is given by
\begin{equation}
h'(0)=\frac{4}{c(0)}=\frac{16\pi}{\Qb{\Pt}{\P}}\cdot\frac{1}{\Delta '}
\end{equation}
and therefore has the same sign as $\Delta'$.  
When $\Delta'>0$, the function $h(\lambdah)$
tends to $-\infty$ for $\lambda$ somewhere between 0 and $\mathrm{min}_{z\in[0,\infty]}(\P)/2$
(namely where $c(\lambda)=0$).
Conseqeuntly, there exists a non--Gaussian FP
with positive $\lambdah_*$.  
On the other hand, when $\Delta'<0$, $h$
has no positive zeroes and the existence of the NGFP for negative
$\lambdah_*$ hinges on the asymptotic behaviour of $h$ for
$\lambdah\to -\infty$: it only exists if $h$ tends to a negative
asymptote.  The asymptotic behaviour of $h$ is given by
$\lim\limits_{\lambda\to -\infty}h(\lambda)=192\pi/\tau$, where
\begin{align}
\label{eq:limith}
\tau=&\tau_0+n_S\tau_S+n_W\tau_W+n_M\tau_M+n_{RS}\tau_{RS}\\
\tau_0=&-5\Qa{\Pt}{\P}-15\Qb{\Pt}{\P^2}+10\Qa{\R}{\P}\notag\\
&+30\Qb{\R}{\P^2}\approx -12.82\notag\\
\tau_S=&2\Qa{\Pt}{\P}\approx 3.58\notag\\
\tau_W=&-4\Qa{\Pt}{\P}+6\Qb{\Pt}{\P^2}\approx -1.62\notag\\
\tau_M=&\Qa{\Pt}{\P}-9\Qb{\Pt}{\P^2}\approx -6.52\notag\\
\tau_{RS}=&16\Qb{\Pt}{\P^2}\approx 14.79.\notag
\end{align}
%
(The numerical values are given for $a=2$).

\begin{figure}\vspace*{3mm}
  \centering{\resizebox{\figurewidth}{!}
    {\includegraphics{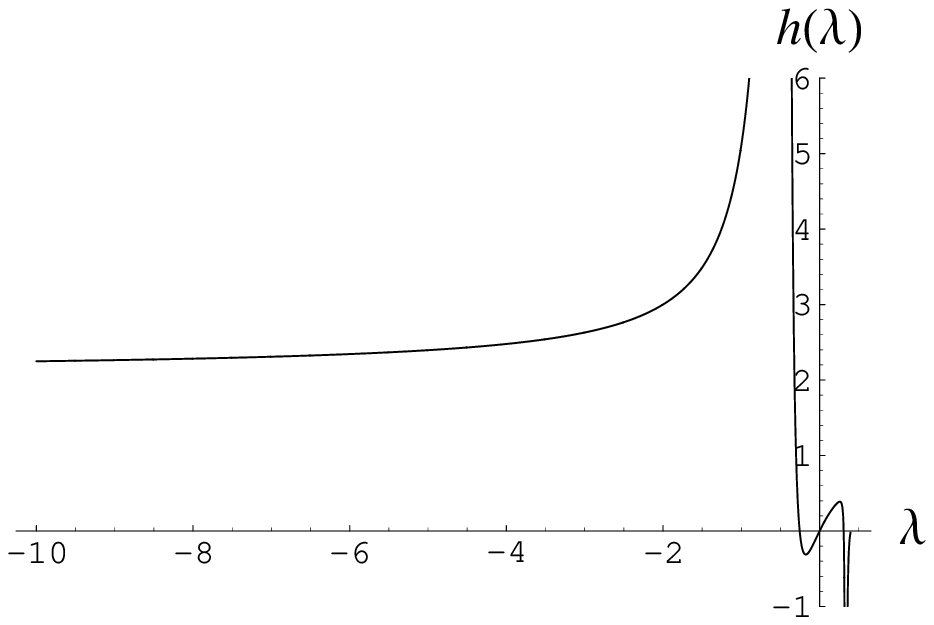}}\\[2mm]
\fbox{\resizebox{.5\figurewidth}{!}
    {\includegraphics{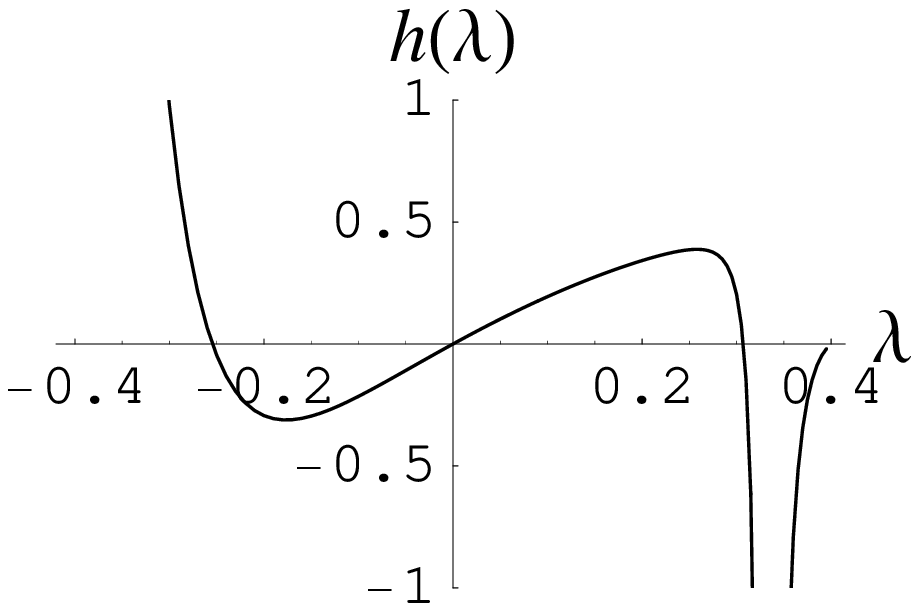}}}
    }
\caption{\label{fig:II.eps} Plot of the function $h$ for
$n_S=100, n_W=40, n_M=0$ (Region II).
The asymptotic value of $h$ for $t\to-\infty$ is 0.939.}%
\end{figure}

\begin{figure}\vspace*{3mm}
  \centering{\resizebox{\figurewidth}{!}
    {\includegraphics{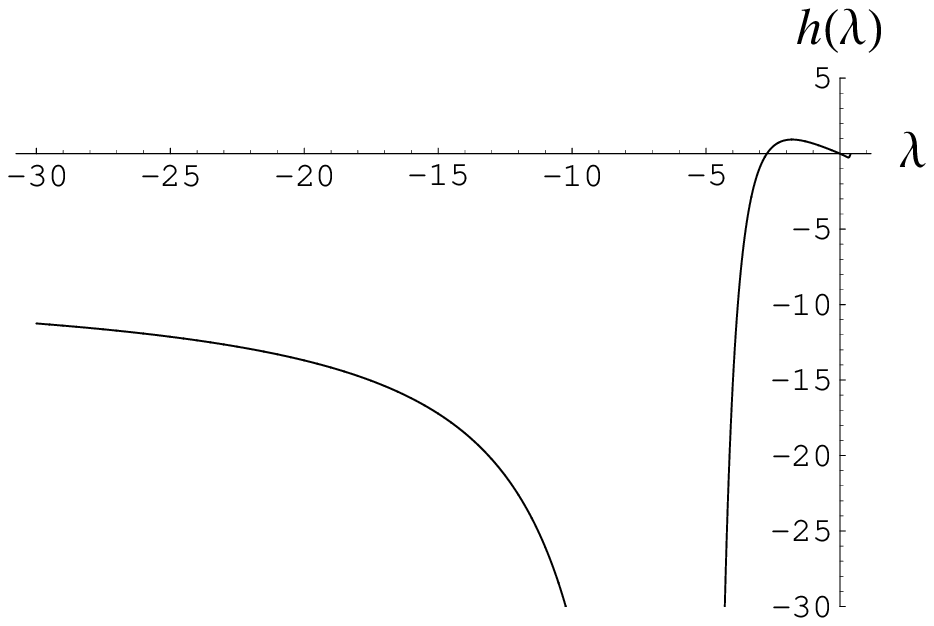}}\\[2mm]
\fbox{\resizebox{.5\figurewidth}{!}
    {\includegraphics{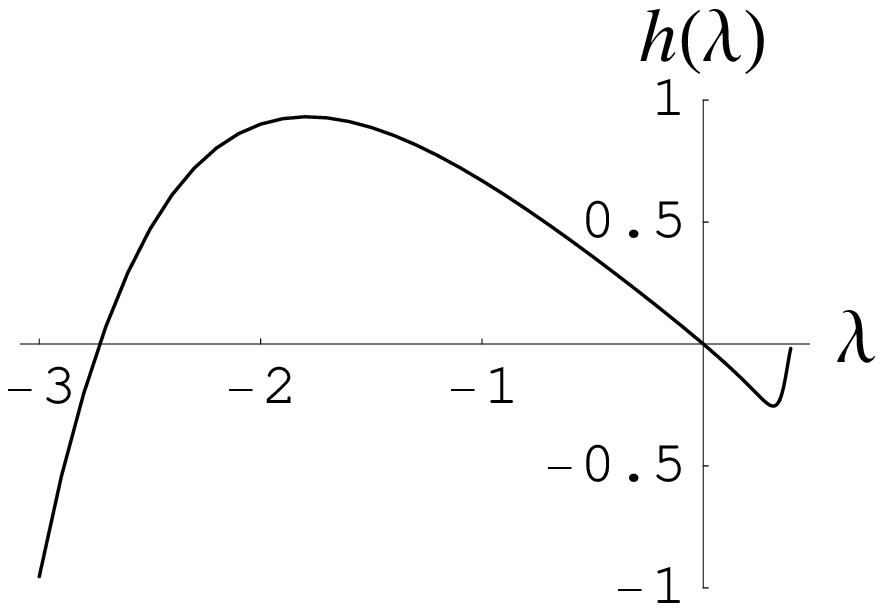}}}
    }
\caption{\label{fig:III.eps} Plot of the function $h$ for
$n_S=0, n_W=40, n_M=0$ (Region III). 
The asymptotic value of $h$ for $t\to-\infty$ is -8.13.}%
\end{figure}
Depending on the sign of the two parameters $\Delta'$ 
and $\tau$, the
space spanned by the variables $n_S$, $n_W$, $n_M$ and $n_{RS}$
can be divided into four regions that we shall label as follows:

\begin{center}
\begin{tabular}{c| c |c|}
&
$\tau<0$&
$\tau>0$\\
\hline
$\Delta'<0$&
III&
IV\\
\hline
$\Delta'>0$&
I&
II\\
\hline
\end{tabular}
\end{center}

The behaviour of the function $h$ is shown in figure 
\eqref{fig:I.eps} for pure gravity, which lies in region I.
There are no zeroes for negative $\lambda$, since $h$ grows
monotonically from the asymptote at $t\to -\infty$
to zero in the origin (Gaussian FP). It then has a positive
zero and tends to $-\infty$.
There is another apparent zero for $\lambda\approx 0.4$,
but it is not an acceptable solution: it corresponds to the
point where the denominators in the function $Q_i$ vanish.
Thus, in region I there is always a single FP
with positive $\lambda_*$.

Figure \eqref{fig:II.eps} shows the function $h$ for a theory
in region II.
The behaviour for positive $\lambda$ is very similar to that in region I,
but the asymptote for $t\to -\infty$ is now positive, so that
there exists a second FP for negative $\lambda_*$
This FP can be seen to yield negative $g_*$, 
and is therefore physically uninteresting.

The behaviour of the function $h$ in region III is shown
in figure \eqref{fig:III.eps}.
The positive zero is the unphysical one, so
there is a single attractive GMFP with 
negative $\lambda_*$, which turns out to have positive $g_*$. 
 
\begin{figure}\vspace*{3mm}
  \centering{\resizebox{\figurewidth}{!}
    {\includegraphics{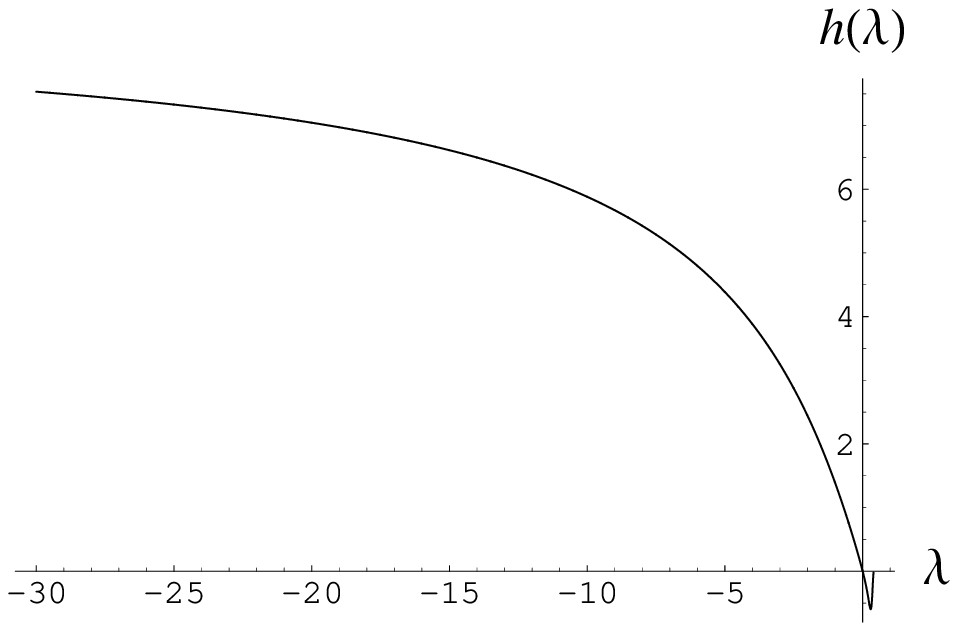}}\\[2mm]
\fbox{\resizebox{.5\figurewidth}{!}
    {\includegraphics{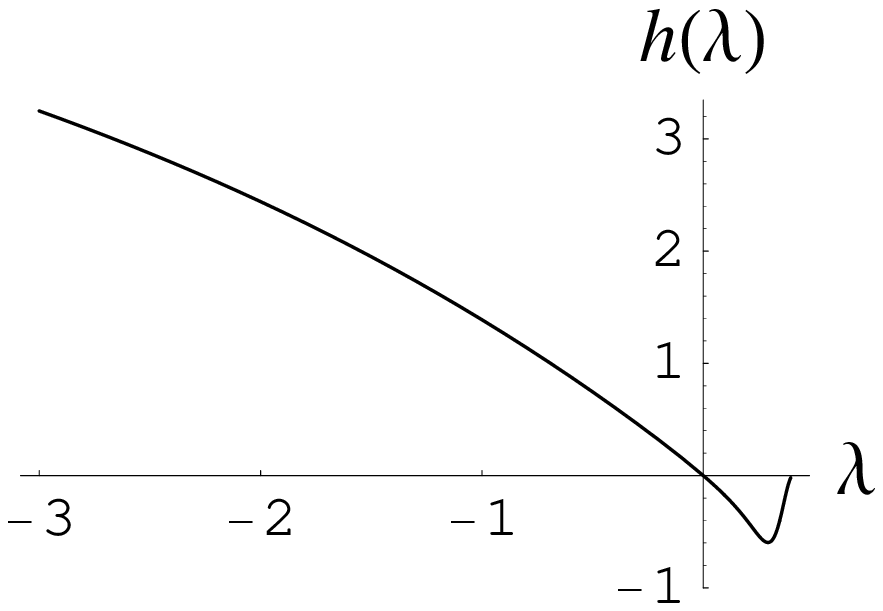}}}
    }
\caption{\label{fig:IV.eps} Plot of the function $h$ for
$n_S=40, n_W=40, n_M=0$ (Region IV). 
The asymptotic value of $h$ for $t\to-\infty$ is 8.72.}%
\end{figure}

\begin{figure}[!t]\vspace*{3mm}
  \centering{\resizebox{\figurewidth}{!}
    {\includegraphics{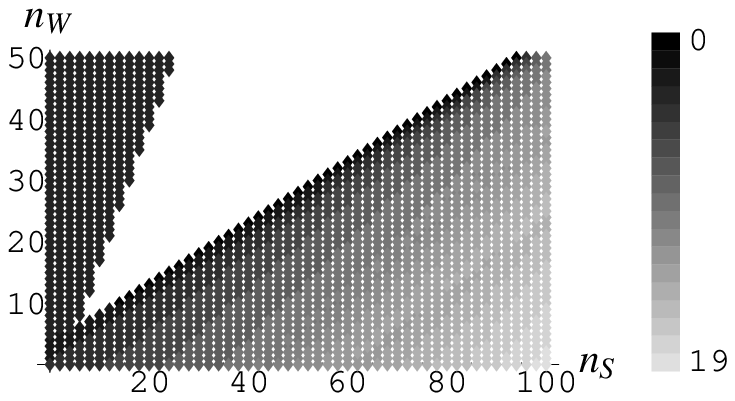}}}
\caption{\label{fig:attr.Nm.0} Attractivity regions for
  $n_M=0$. The colours correspond to the number of
  attractive directions.}
\end{figure}

\begin{figure}[!t]\vspace*{3mm}
  \centering{\resizebox{\figurewidth}{!}
    {\includegraphics{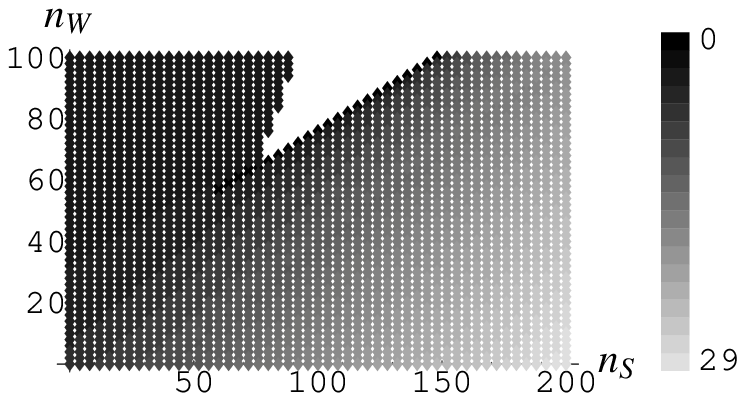}}}
\caption{\label{fig:attr.Nm.24} Attractivity regions for $n_M=24$.
  The colours correspond to the number of attractive directions.}
\end{figure}

\begin{figure}[!t]\vspace*{3mm}
  \centering{\resizebox{\figurewidth}{!}
    {\includegraphics{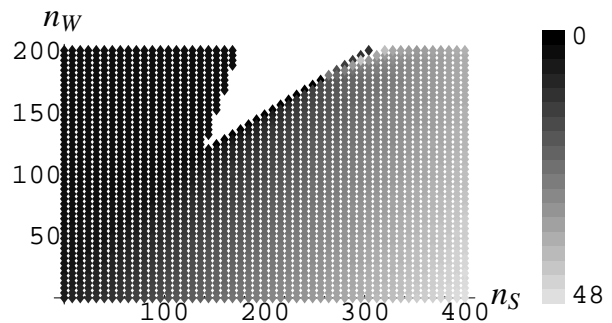}}}
\caption{\label{fig:attr.Nm.45} Attractivity regions for $n_M=45$.
   The colours correspond to the number of
  attractive directions.}%
\end{figure}

Finally, the behaviour of the function $h$ in region IV is shown
in figure \eqref{fig:IV.eps}.
It decreases monotonically from the positive asymptote
$t\to-\infty$ to the Gaussian FP.
For positive $\lambda$ it behaves like in region III,
having no zeroes except for the unphysical one.
Thus, in region IV there is no non-Gaussian FP.
Region IV is the white wedge in figures
(\ref{fig:attr.Nm.0},\ref{fig:attr.Nm.24},\ref{fig:attr.Nm.45}).
One sees that it comes actually quite close to the origin;
from this point of view the existence of the FP for pure gravity 
seems to be a lucky accident.

The value of $\lambda_*$ in regions I and II is always less than
$\mathrm{min}_{z\in[0,\infty]}(\P)/2$, which is numerically
equal to 0.402 (for $a=2$) and
therefore reasonably within the bounds of the heat-kernel
approximation. On the other hand in region III $\lambda_*$ becomes
quickly rather large in absolute value; in this regime $R\gg k^2$ on
shell and therefore the heat-kernel approximation ceases to be valid
on shell.
In this region the results are only reliable close to the surface
$\Delta'=0$.

In order to determine the dimension of the critical surface we have
calculated numerically the matrix $M$ for many different combinations
of fields. The results of such calculations are shown in figures
(\ref{fig:attr.Nm.0},\ref{fig:attr.Nm.24},\ref{fig:attr.Nm.45})
for the case $n_{RS}=0$ and $n_M=0$, $n_M=24$ and $n_M=45$ respectively
(these numbers are chosen
to correspond to the gauge field content of popular GUT models).

First of all, these numerical calculations exactly confirm the shape
of the existence region of the FP that was derived analytically above.
The structure of the eigenvalues is the same as in the pure
gravity+scalar case, which was discussed in Section IV.  The
eigenvalues are given by the canonical dimensions plus a quantum
correction which depends on the type and number of matter fields but
otherwise is the same for every pair $(\l{2i},\x{2i})$.  For any given
number of matter fields, the GMFP has a finite-dimensional critical surface.
In region III, the critical surface has mostly dimension 3 
(three negative real eigenvalues),
except for a narrow area close to the separatrix
$\Delta'=0$, where its dimension is 2 or 4.  In regions I and II the
critical dimension varies considerably, being roughly linear in the
number of fields (it grows with $n_S$ and decreases with $n_W$).
These calculations generalize the results of \cite{Percacci:2002ie},
where the FP could have at most 2 attractive directions.

It is interesting to compare these results with the analysis of
the pure scalar theory. Due to the fact that the coupling $\l4$
is marginal in the pure scalar theory, the linearized analysis
is not sufficient to determine its behaviour.
In the presence of gravity there is no zero eigenvalue and therefore
the linearized analysis is sufficient to determine the dimension
of the critical surface.
In region III, the relevant directions correspond to potentials
that are at most quadratic in $\phi$.
In region II, however, there can be a large number of negative eigenvalues,
corresponding nontrivial potentials $V$ that are
polynomial and asymptotically free, thus avoiding entirely
the triviality issue.
These theories are also predictive 
since they have a finite number of negative eigenvalues
\footnote{This should be contrasted with the asymptotically free
nonpolynomial potentials of \cite{Halpern:1995vw}.
Those potentials are parametrized by a continuous parameter
and therefore have an infinity of relevant couplings.}.
They are therefore asymptotically safe.

\section{Cutoff and gauge dependence}
Physical results are independent of cutoff parameters in the exact
theory, so the extent of parameter dependence that is observed in the
truncated theory gives a quantitative measure of the errors.  
We have
performed various tests on the parameter dependence of our results and
it is reassuring for the reliability of the truncation that this
dependence turns out to be reasonably mild. 

\begin{figure}[t]
    \center{\resizebox{\figurewidth}{!}{\includegraphics{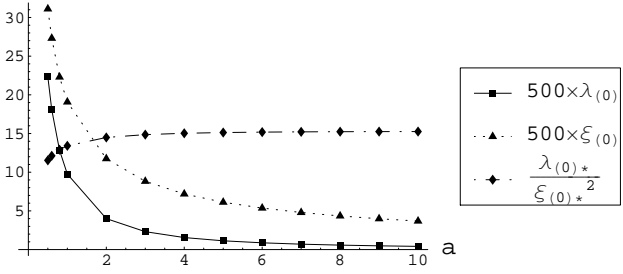}}
    \caption{\label{fig:cutoff_indep}$a$-dependence of
      $\l{0*}$, $\x{0*}$ and $\l{0*}/\x{0*}^2$
      in the gauge $\alpha=0$. The values of $\l{0*}$ and
      $\x{0*}$ have been magnified by a factor 500 to display the
      three curves in the same range.}}
\end{figure}

The dependence of $\lambda_{0*}$ and $\xi_{0*}$ on gauge and
cutoff parameters was discussed in \cite{Lauscher:2001ya,Souma:2000vs}.
Figure~(\ref{fig:cutoff_indep}) summarizes the cutoff-dependence at
the GMFP for gravity coupled to one scalar field in the gauge
$\alpha=0$. The results we obtain are very close to those of
\cite{Lauscher:2001ya}, since at the GMFP the only new contribution
that we get for the values of $\lambda_{0*}$ and $\xi_{0*}$ is
that of the kinetic term of the scalar field. It is apparent that
while $\lambda_{0*}$ and $\xi_{0*}$ are quite sensitive to the
cutoff parameter $a$, the ratio $\lambda_{0*}/\xi_{0*}^2$ is not. 
As noted in \cite{Dou:1998fg}, this quantity is, up to numerical factors,
(the inverse of) the on-shell action, a physically observable quantity, 
so it must be independent of the cutoff scheme. 
It is seen in figure~(\ref{fig:cutoff_indep}) that its
$a$-dependence is indeed pretty mild. 

\begin{figure}[!t]
    \center{\resizebox{\figurewidth}{!}{\includegraphics{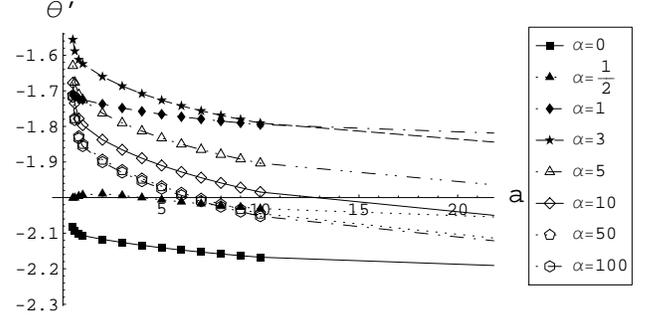}}
    \caption{\label{fig:eigenre.1}$a$-dependence of the real part of
      the eigenvalues of the stability matrix (first $2\times 2$ submatrix).}}
\end{figure}

\begin{figure}
    \center{\resizebox{\figurewidth}{!}{\includegraphics{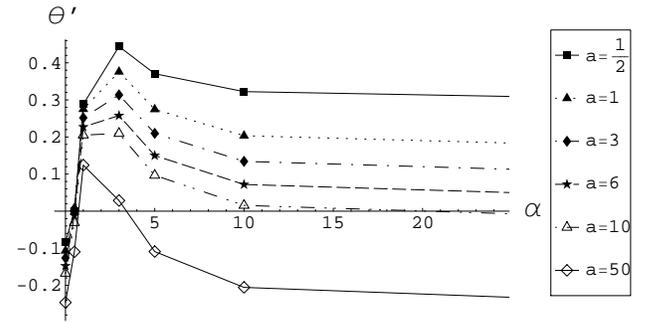}}
    \caption{\label{fig:eigenre.alpha}$\alpha$-dependence of the real part
      of the eigenvalues of the stability matrix (second $2\times 2$
      submatrix).}}
\end{figure}

\begin{figure}[h!]
    \center{\resizebox{\figurewidth}{!}{\includegraphics{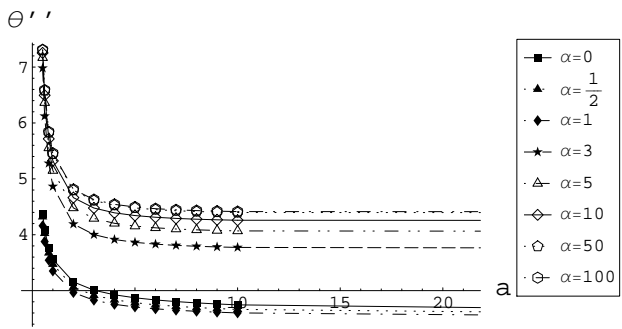}}
    \caption{\label{fig:eigenim}$a$-dependence of the imaginary part
      of the eigenvalues of the stability matrix.}}
\end{figure}

The dependence of 
$\theta'_{2i}\pm i\theta''_{2i}$,
the eigenvalues of the stability matrix $M$,
on the cutoff parameter $a$ is shown in
figures~(\ref{fig:eigenre.1}-\ref{fig:eigenim}), for several values of
$\alpha$. We have calculated them in the range $1/5\leq a\leq50$, but
they are only reported for $1/2\leq a\lesssim20$.
Figure~(\ref{fig:eigenre.1}), giving the real parts of the eigenvalues
of the submatrix $M_{00}$, agrees with figure~(9) of
\cite{Lauscher:2001ya}, up to the small corrections due to the
presence of a scalar field. The figures relative to real parts of the
remaining eigenvalues are simply shifted by the canonical
dimension $2i$.

The first thing we can notice in figure (\ref{fig:eigenre.1}) is the
presence a clear plateau with very weak (apparently logarithmic)
variation of the eigenvalues, for $1\lesssim a\lesssim20$. Actually,
the results for $1/5\leq a\leq1/2$ seem to indicate that there is a
divergence as $a\to 0$. 
This is due to the fact that in this limit the
cutoff function tends to become a constant, 
so it affects also the propagation of modes with momenta
larger than $k$, and it does not work well as in IR cutoff. 
Clearly, larger values of $a$, of order unity, are preferred.
This is in accordance with the
generic features of the cutoff functions described in
\cite{Wetterich:1993yh}.
%
%
%
%

As far as the $\alpha$ dependence is concerned, we can
notice that it is quite weak. For all
possible values of $\alpha$ all curves are contained
between the curves $\alpha=0$ and $\alpha\simeq3$, which differ by
$\sim 0.4$. In order to better understand the dependence on $\alpha$
for different values of $a$, it is useful to plot the same results as
a function of $\alpha$ (figure (\ref{fig:eigenre.alpha})), with $a$
being a parameter that labels different curves (we shall restrict only
to one plot of the real parts, the others can be obviously derived by
shifting the graph by the canonical dimension of the operator
involved).

Since the real parts of the eigenvalues of the matrix $M_{22}$ are
close to zero, this modest shift of the eigenvalues due to the change
of gauge parameter is enough to change their sign.  For instance, in
the gauge $\alpha=1$ one would compute the dimension of the
critical surface to be two. Looking at figure
(\ref{fig:eigenre.alpha}) one can most easily understand what the
situation is like: for $\alpha=0$ all cutoffs give a negative value of
$\theta'$, then as $a$ increases they change sign, but for large
values of $\alpha$ they become negative again. Physical results such
as the dimension of the critical surface cannot depend on the shape
of the cutoff function,
so this fact is certainly a shortcoming of our truncation.  
More work is needed to assess with greater confidence the dimension of the 
critical surface, but the considerations developed in 
\cite{Lauscher:2001ya}, \emph{i.~e.}\
that $\alpha$ itself runs to 0 in the UV regime, suggest that the
$\alpha=0$ value is the physicaly correct result.

The same conclusions for the cutoff and gauge indepence can be drawn
for the imaginary parts, as can be seen from figure
(\ref{fig:eigenim});%
they turn out to take the same values (up to a
sign) for all the eigenvalues. The effect of nonvanishing imaginary parts
is that the RG spirals around the FP, but they are not
important in the discussion of the attractivity of the FP.

This discussion applies also to the higher couplings;
their $a$- and $\alpha$-dependence is given by curves that
differ from those in figures (\ref{fig:eigenre.1}--\ref{fig:eigenim})
by a constant shift by a multiple of 2.
The only important point that remains to some extent open, then, is
the exact dimension of the UV critical surface, but
nevertheless we can safely say that it is finite-dimensional.

When other matter fields are present, the nature of the GMFP as a
function of the number af matter fields is also a function of gauge
and cutoff parameters.  As already noted in \cite{Percacci:2002ie},
the constant $\sigma$ is independent of the gauge parameter and varies
from 4.745 for $a=0.05$ to 2.765 for $a=20$. This corresponds
to a vertical shift of the separatrix $\Delta'=0$ by at most 2
in figures (\ref{fig:attr.Nm.0}--\ref{fig:attr.Nm.45}).
The parameter $\tau$ is
gauge independent for $\alpha\neq0$, but shows discontinuity at
$\alpha=0$. 
The plane $\tau=0$ is shifted and also slightly rotated to the right as
$a$ grows.  Thus, region III becomes larger as $a$ grows.  
Recall however that only the part of this region close to the
separatrix $\Delta'=0$ is trustworthy.


\section{Conclusions}
In this paper we have considered the application of the ERGEs to a coupled system
 of gravity and matter fields. The
main aim of this work was to verify that the conditions for asymptotic safety
continue to hold in the presence of interacting matter fields.
To make the problem manageable, we have first dealt with a single
scalar field $\phi$ with an arbitrary potential depending on $\phi^2$,
to see how this inclusion could change the picture of pure gravity,
then we have considered the effect of minimally coupled fields with
different spins. Our results can be considered as a first step towards
constructing a realistic theory of gravity and matter, but are also 
relevant to gravitational theories containing a dilaton.

In the context of the Ansatz \eqref{eq:class} we found that 
there exists a FP where only the cosmological constant 
and Newton's constant are nonzero. 
We called it the ``Gaussian Matter'' FP.
A detailed numerical search within a five-parameter truncation 
of the effective action has failed to yield any other FP. 
This is actually what one would expect from our understanding
of the scalar theory \cite{Halpern:1995vw}.

The GMFP may be viewed in two ways.  On one hand, the scalar field can
be regarded as a ``perturbation'' of the pure theory of gravity
considered in \cite{Lauscher:2001ya} and the GMFP as an extension of
the FP found in \cite{Souma:1999at}.  The addition of the scalar field
has the effect of shifting slightly the values of $\Lambda$ and
$\kappa$, as already noticed in \cite{Percacci:2002ie}, while the
attractivity property is preserved.  On the other hand, we can consider
the effect of adding gravity to a scalar theory with a generic
potential and regard the GMFP as an extension of the Gaussian FP.  The main
effect is that the couplings that are present in the scalar potential
$V$ mix with those appearing in the function $F$, and the dimensions
of the resulting couplings (which dictate the speed of the approach to
the FP) is changed by a finite quantum correction.  While at the
Gaussian FP the quantum corrections vanish, so that the relevant
couplings are, as usual, those with dimension less than four, the
gravitational contributions bring about modifications even if the
matter sector allows for a perturbative treatment.  At the GMFP, the
stability matrix has a block-diagonal form so that there is a strong
ing between the parameters $\l{2n}$ and $\x{2m}$ for $n=m$,
whereas for $n\neq m$ they are almost or completely decoupled. The
eigenvalues, whose real part determines whether an operator is
relevant or irrelevant, come in complex-conjugate pairs, and grow
systematically by a constant 2. For instance, the marginal operator of
the pure-scalar $\phi^4$ theory becomes now an irrelevant operator, 
and the dimension of the UV critical surface is
calculated to be 4 for a generic analytic potential.
These results hold in the gauge $\alpha=0$; they differ slightly
for other values of the gauge-fixing parameter, but $\alpha=0$ seems
to be the physically correct value at the FP.
The striking fact is that
gravity gives calculable, finite contributions that change
significantly the pure-scalar theory.  This is one of the most
important results of our paper.

We have then considered the effect of adding other minimally
coupled massless matter fields.
For the existence of the GMFP, we obtain
the same bounds presented in \cite{Percacci:2002ie}. As to the
attractivity of the FP, we have found that, when it exists, there are
always finitely many attractive directions. Therefore gravity seems to remain
asymptotically safe also in the presence of generic matter fields.
We expect that this result will still hold if we add other interactions
between matter fields that are asymptotically free.
From this point of view, the scalar field posed a greater challenge,
since the pure scalar theory is not asymptotically free.
It is remarkable that the coupling to gravity fixes this problem
and at the same time also offers a solution to the triviality problem.

All these results add on to the other proofs
that have been collected in the literature about
the physical reliability of this approach. %
\begin{acknowledgments}
We thank A.~Bonanno and  M.~Reuter for useful discussions.
\end{acknowledgments}

\bibliographystyle{apsrev}
\bibliography{database}

\end{document}